\documentstyle[epsf]{article}

  \font\tenbmi=cmmib10 at 10pt  \skewchar\tenbmi ='177
  \font\sevenbmi=cmmib10 at 10pt \skewchar\sevenbmi ='177
  \font\fivebmi=cmmib10 at 10pt  \skewchar\fivebmi ='177

  \newfam\bmifam
  \textfont\bmifam=\tenbmi
  \scriptfont\bmifam=\sevenbmi
  \scriptscriptfont\bmifam=\fivebmi
  \def\bmi{\fam\bmifam\tenbmi}
\begin{document}

\title {Can we look inside a dynamo?}

\author{Frank Stefani, Gunter Gerbeth\\
Forschungszentrum Rossendorf, Germany\\
P.O. Box 510119, D-01314 Dresden, Germany}
 
\maketitle

%%%%%\summary
%%%%% Text of summaryEND
\begin{abstract}
For a simple spherically symmetric 
mean-field dynamo model we investigate the possibility of 
determining the radial dependence 
of the coefficient $\alpha$. Growth rates 
for different magnetic field modes 
are assumed
to  be known by measurement. 
An evolutionary strategy (ES) is used for the solution of the
inverse problem. Numerically, we find quite different $\alpha$-profiles 
giving nearly the same eigenvalues. The ES is also applied to find
functions $\alpha(r)$ yielding zero growth rates for the
lowest four magnetic field modes. Additionally, a slight modification 
of the ES is utilized for an ''energetic'' optimization
of $\alpha^2$-dynamos. 
The consequences of our findings
for inverse dynamo theory and for the design of future 
dynamo experiments are discussed.
\end{abstract}

\section{Introduction}
At the very beginning, the question of the origin of 
cosmic magnetic fields was
formulated as an inverse problem. ''How could a rotating
body such as the sun  become a magnet?'' was the question
Larmor put in 1919 (Larmor 1919), i.e. he asked for the unknown
cause of an evident effect. In the last decades, dynamo theory
saw an enormous progress. Testing a large number of 
kinematic models in the sense of a forward problem, the inverse
character of the original problem was partly  pushed into
the background. It seems to be only a few percent
of  papers on the dynamo topic which are 
dealing with an inverse problem 
in the strict sense. 

In geophysics, there is a long tradition
to reconstruct the tangential components 
of the velocity field at the core-mantle boundary from
geomagnetic secular variations
 (see, e.g, Bloxham (1989), and
references therein). But the value 
of this type of inverse frozen-flux modeling  
was called into question recently (Love 1999). The 
main argument of the criticism was that the 
removal of the diffusion term from the induction equation
changes its differential order resulting possibly in a
drastic change of the nature of the solutions.

Another approach to the inverse dynamo problem was
formulated by Love and  Gubbins (1996). In this paper, 
the eigenvalue 
equation was complemented by some regularizing 
functionals in order to optimize dynamo models 
with respect to energetic demands and smoothness properties
of the arising magnetic field. However, geomagnetic 
observations 
were taken into account only in the sense that some
smoothness properties of the model magnetic field were 
compared with 
those of the observed field and not in the sense of 
fitting the model 
to the actual data. 
  
With the advent of the first successful hydromagnetic 
dynamo experiments (Gailitis et al. 2000; M\"uller and 
Stieglitz 2000) 
quite
new perspectives for inverse dynamo theory 
come into play. Compared to cosmic bodies, much more 
measurement techniques and measurable  
quantities can be taken into 
account. One additional measurable 
quantity is the electric potential
at the boundary of the dynamo module which can provide 
complementary information  in addition to the magnetic field
on the outside. Further on, in contrast to the case of cosmic 
bodies it
is possible to apply various magnetic fields from outside 
and to determine the response of the dynamo on these primary 
fields.
Those date can be collected in the regime below
the critical magnetic Reynolds number ($Rm$) in order 
to predict
the dynamo behaviour at larger $Rm$. 

In connection with this, 
but up to now restricted to small $Rm$,
there are some new developments concerning
the determination of the velocity field from 
externally measured 
induced magnetic fields and electric potentials when
primary magnetic fields are applied (Stefani and 
Gerbeth 1999, 2000a, 2000b).
For a static homogeneous primary field and 
spherical geometry it was shown 
(Stefani and Gerbeth 2000a) that the defining scalars of 
the velocity 
can be determined except for an ambiguity of the radial 
dependence of their spherical harmonics expansion
coefficients. Although one can try to get a reasonable 
guess of the unknown radial dependence by some
regularization techniques, in the strict sense this dependence
remains unobservable. A generalization of this method 
to higher $Rm$ would amount to a combination of an 
eigenvalue solver (similar to the philosophy of Love and Gubbins)
with a data-fitting procedure. However this approach might 
look like in detail, we guess that the radial dependence 
will remain undetectable. Roughly speaking, two
two-dimensional
measured quantities (radial component of magnetic field plus 
electric potential) allow only to determine two
two-dimensional desired quantities. Actually, this 
radial ambiguity was the first reason 
for the following investigation. 
At least for the small $Rm$ regime 
it seems clear that the necessary third dimension to 
determine
the radial dependence can only be the time (or frequency).

Whereas a general inverse dynamo theory is 
still missing, it 
is worth to have a sidelong glance at the 
quite similar
inverse problem in quantum mechanics. There is
a huge literature on inverse scattering theory and
inverse spectral theory (for an overview, see 
Chadan and Sabatier (1977) and Chadan et al. (1997)), and 
evidently many methods  from quantum mechanics 
(Gelfand-Levitan equation, WKB methods, operator 
methods, etc.)
remain to be adapted and utilized  in inverse 
dynamo theory.  

In the present paper, we will 
pursue a less ambitious program 
restricting ourselves to a pragmatic treatment
of a simple inverse spectral dynamo problem.
It is a simple problem as it concerns 
spherically symmetric $\alpha^2$-dynamos 
in which, as a slight modification of the 
original  model of Krause and Steenbeck (1967),
the radial dependence of $\alpha$ is assumed 
to be unknown.
The treatment will be pragmatic
in the sense that we suppose  
eigenvalues of only a few magnetic field
modes
to be "measured" and
that we will try to infer the radial dependence
of $\alpha$ from this restricted set of information.

Basically,  it is our aim to get some acquaintance with 
the peculiarities if one tries to get information 
on the radial 
distribution of the source of dynamo action. Reminding 
quantum theory might be useful  in the following 
respect: The existence of 
isospectral 
potentials  is a well established fact
in quantum theory. For example, it is possible to 
deform {\it{continuously}} the harmonic oscillator 
potential in such a way that all the 
eigenvalues are the same as for the original 
quadratic potential (De Lange and Raab 1991). 
Therefore, we should be aware of the possibility 
that any
inversion procedure of any set 
(even of an infinite number) 
of measured eigenvalues might lead to a variety of
radial $\alpha$-distributions
which cannot be discriminated using only 
their eigenvalues. For the 
corresponding quantum inverse problem it is known that
a potential can be uniquely determined if 
the spectra {\it{for two different boundary conditions}} 
were known. A similar behaviour should also be expected 
for the inverse dynamo problem, but this
seems to be irrelevant for any realistic 
measurement strategy.

Our method might have application for cosmic
dynamos as well as for laboratory dynamos. 
Of course, a spherically symmetric $\alpha^2$-dynamo 
model 
is surely  too simple to explain 
cosmic dynamos but the general method
is easily extendable to more realistic  models.

The paper is structured as follows: in the second section, the 
used eigenvalue solver is presented and
five paradigmatic types of radial dependence of $\alpha$ 
are treated
as a forward problem. 
In the third section, the evolutionary strategy  to invert the
''measured'' growth rates for a number of magnetic field 
modes into a radial
dependence of $\alpha$ is presented. Then, we try to reconstruct 
five models from section 3. Numerically, we find 
quite different $\alpha$-profiles 
giving nearly the same eigenvalues.
Additionally,  we treat two other problems which can be, 
at least in 
principle, of interest for future dynamo experiments. 
At first we try to find such a function $\alpha(r)$ which yields
zero growth rate for the lowest four field modes. 
Such a profile for $\alpha$ might be a candidate for a 
(hypothetical) 
dynamo 
experiment showing such interesting effects like 
mode-switching.  
Further on, we use our code for an ''energetic'' optimization
of $\alpha^2$-dynamos.
 The paper closes with
some  remarks on possible  future developments.

\section{The forward problem}
Shortly speaking, our approach to solve the inverse problem 
will be the most simple one: to solve the forward
problem many times and to seek, in some appropriate manner,
 for those 
dynamo models which give eigenvalues as close as possible 
to the measured ones. Therefore, the present chapter on the
forward problem
describes also the kernel of the inverse problem solving code 
which will be presented in detail in the next section. 

We start with the induction equation 
for a  mean-field dynamo model restricted to a 
spherically symmetric
$\alpha$-coefficient.  For the sake of simplicity, 
we ignore any large-scale velocity 
field as well as any $\beta$-effect or any anisotropic
$\alpha$-coefficients. The electrical conductivity $\sigma$ is
assumed to be constant inside a sphere of radius $R$ 
and is assumed to be
zero
in the outer part. Then, inside the sphere, the 
magnetic field has to satisfy the equations
(Krause and R\"adler 1980)
\begin{eqnarray}
\frac{\partial {\bmi{B}}}{\partial t} =\nabla 
\times (\alpha {\bmi{B}}) +
\frac{1}{\mu_0 \sigma} \Delta {\bmi{B}} \; , 
\;\;\;  \nabla \cdot {\bmi{B}}=0 \; .
\end{eqnarray}
At the boundary with $r=R$, the magnetic field has to match 
continuously to a 
potential field. 

In the usual manner, $\bf{B}$ is represented as a sum of
poloidal and toroidal components,
\begin{eqnarray}
{\bmi{B}}=-\nabla \times ({\bmi{r}} \times 
\nabla S)-{\bmi{r}} \times 
\nabla T
 \end{eqnarray}
with the defining scalars $S$ and $T$ expanded 
in spherical harmonics
according to 
\begin{eqnarray}
S(r,\theta,\phi)&=&\sum_{l=1}^{\infty} \sum_{m=-l}^{l}
R \, S_l^m(r) Y_l^m(\theta,\phi) \exp{\lambda_{l} t}\\
T(r,\theta,\phi)&=&\sum_{l=1}^{\infty} \sum_{m=-l}^{l}
T_l^m(r) Y_l^m(\theta,\phi) \exp{\lambda_{l} t} \; .
\end{eqnarray}
In order to simplify the notation, throughout the rest 
of the paper we will measure 
the length in units of $R$, the time in units of 
$\mu_0 \sigma R^2$, and
the coefficient $\alpha$ in units of $(\mu_0 \sigma R)^{-1}$.

Using (2), (3) and (4), the induction equation 
can be transformed into the following form (R\"adler 1986):
\begin{eqnarray}
\lambda_l S_l&=&
\frac{1}{r}\frac{d^2}{d r^2}(r S_l)-\frac{l(l+1)}{r^2} S_l
+\alpha(r) T_l\\
\lambda_l T_l&=&
\frac{1}{r}\frac{d}{dr}\left( \frac{d}{dr}(r T_l)-\alpha(r)
\frac{d}{dr}(rS_l) \right) -\frac{l(l+1)}{r^2} 
(T_l-\alpha(r)
S_l) \; .
\end{eqnarray}
Note that there is no coupling between field modes differing in 
the order $l$ of the spherical harmonics.
Furthermore, we have skipped the index $m$ for the coefficients 
of the defining scalars
as it does not show up in the
equations.
Note in addition that
the eigenvalues $\lambda$ may be complex.

At the surface $r=1$, the boundary conditions
\begin{eqnarray}
\frac{d S_l}{dr}+{(l+1)} S_l=T_l=0
\end{eqnarray} 
must be fulfilled.

In order to have a comparable global measure of the
$\alpha$-effect we introduce the following
definition:
\begin{eqnarray}
C=3 \int_0^1 \alpha(r) r^2 dr \; .
\end{eqnarray}
Of course, other definitions, e.g. one without the $r^2$-term 
in the integral,
are as well conceivable. The definition used here 
might be sensible 
for the purpose of ''energetic'' optimization as it reflects the
volume averaged intensity of $\alpha$.

To solve the coupled system of equations (5) and  (6) together 
with the
boundary condition (7) we have applied a standard 
shooting procedure using
Newton's method. 
This code was validated for the case of constant
$\alpha$ as well as for some known results with a 
radial variation
of $\alpha$ (R\"adler 1986).

In the following, we will present the results of the 
forward problem for five paradigmatic models.
Table 1 shows the corresponding functions
explicitely.
The somewhat strange numerical factors arise due to 
the definition (8)
The examples a, b, c, d, and e represent 
the following
types of $\alpha$-profiles: 
type a1 gives a constant $\alpha$ whereas 
$a2$ is only a slightly modified version with a small 
oscillation around 
the constant value. Curves of type b and c are 
monotonically increasing 
and decreasing, respectively.
Curves of type d and e have one maximum and one 
minimum between 0 and 1, respectively.

\begin{table}[t]
\caption{Functions $\alpha(r)$ for ten considered models.}
\begin{center}
\begin{tabular}{llll}
\hline
Model&$\alpha(r)$&Model&$\alpha(r)$\\
\hline
a1&$C$&a2&$C \;(1-5 r^2+12r^3-7r^4)$\\
b1&$5/3 \;C\; r^2$&b2&$7/3 \; C \;r^4$\\
c1&$15/6 \;C \;(1-r^2)$&c2&$21/12\; C \;(1-r^4)$\\
d1&$10 \;C \;(r^2-r^3)$&d2&$35/6\; C \;(r^2-r^4)$\\
e1&$270/13 \;C \;(4/27-r^2+r^3)$&e2&$140/11 \;C \;(1/4-r^2+r^4)$\\
\hline
\end{tabular}
\end{center}
\end{table}

In Fig. 1 the dependence of the growth rates $\lambda$ on 
$C$ for the modes with $1 \le l \le 6$ is shown.
For the sake of simplicity, we take 
into account only the lowest radial wavenumber $n=1$. 
Admittedly, it is not completely consistent 
to include, e.g.,  the mode with $l=3$ and $n=1$ having 
a growth rate of $\lambda(C=0)=-33.1$ and 
to skip 
at the same time the mode with $l=1$ and $n=2$ 
having a growth rate 
of $\lambda(C=0)=-20.2$ (Krause and R\"adler 1980). 
In principle, there is no
obstacle to take into account
 modes with higher radial wavenumbers. 
The restriction to $n=1$ might suffice for the 
simple toy model of inverse dynamo theory as it is 
considered here.  

\begin{figure}[t]
\begin{center}
\begin{tabular}{cccc}
\raisebox{35mm}{a1}&
\hspace{-1.cm}\epsfxsize=6cm\epsfbox{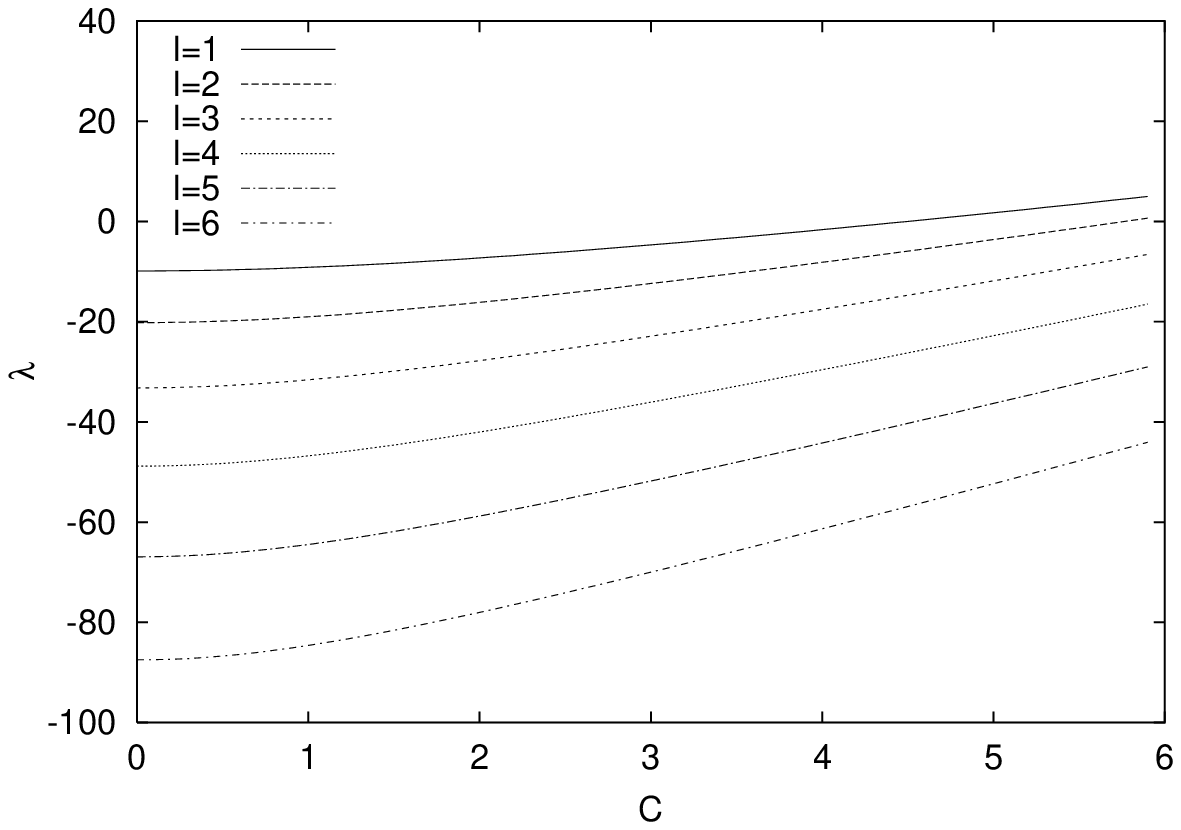} 
&\hspace{1.cm}\raisebox{35mm}{a2}&
\hspace{-1.cm} \epsfxsize=6cm\epsfbox{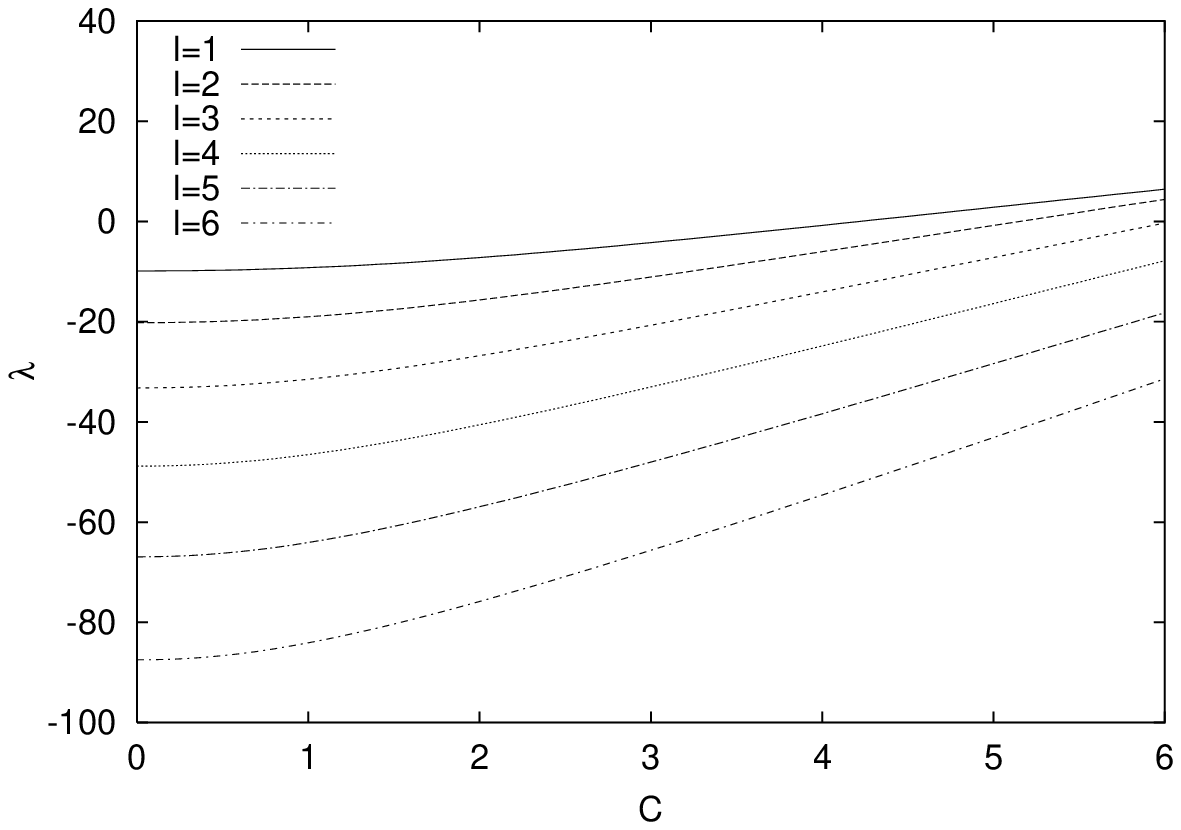}\\
\raisebox{35mm}{b1}&
\hspace{-1.cm}\epsfxsize=6cm\epsfbox{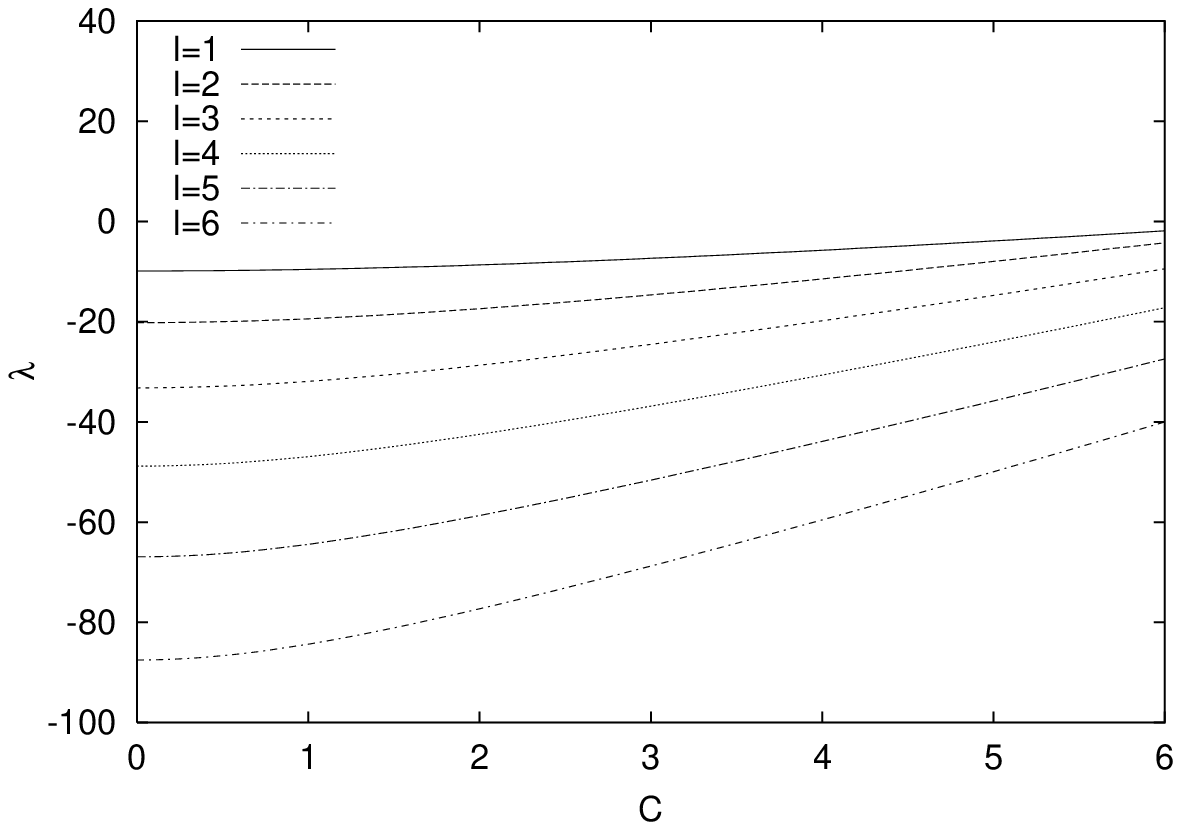} 
&\hspace{1.cm}\raisebox{35mm}{b2}&
\hspace{-1.cm} \epsfxsize=6cm\epsfbox{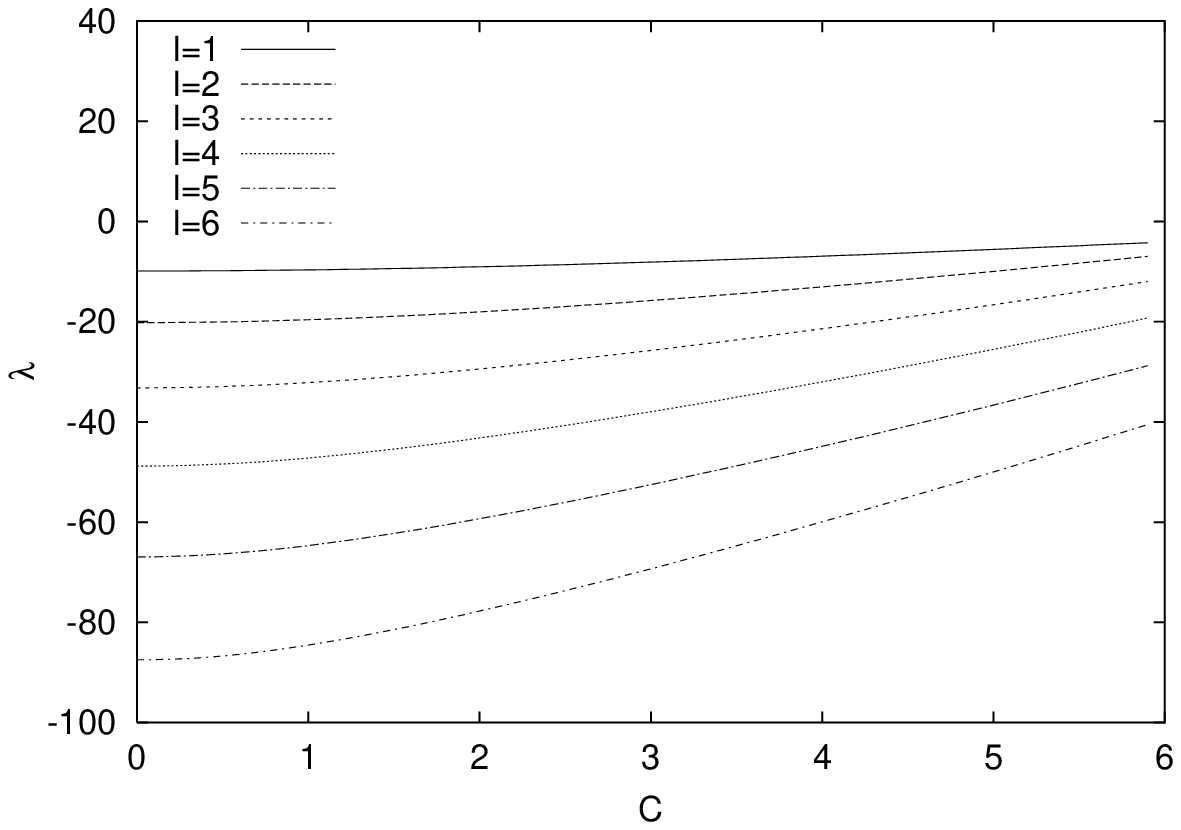}\\
\raisebox{35mm}{c1}&
\hspace{-1.cm}\epsfxsize=6cm\epsfbox{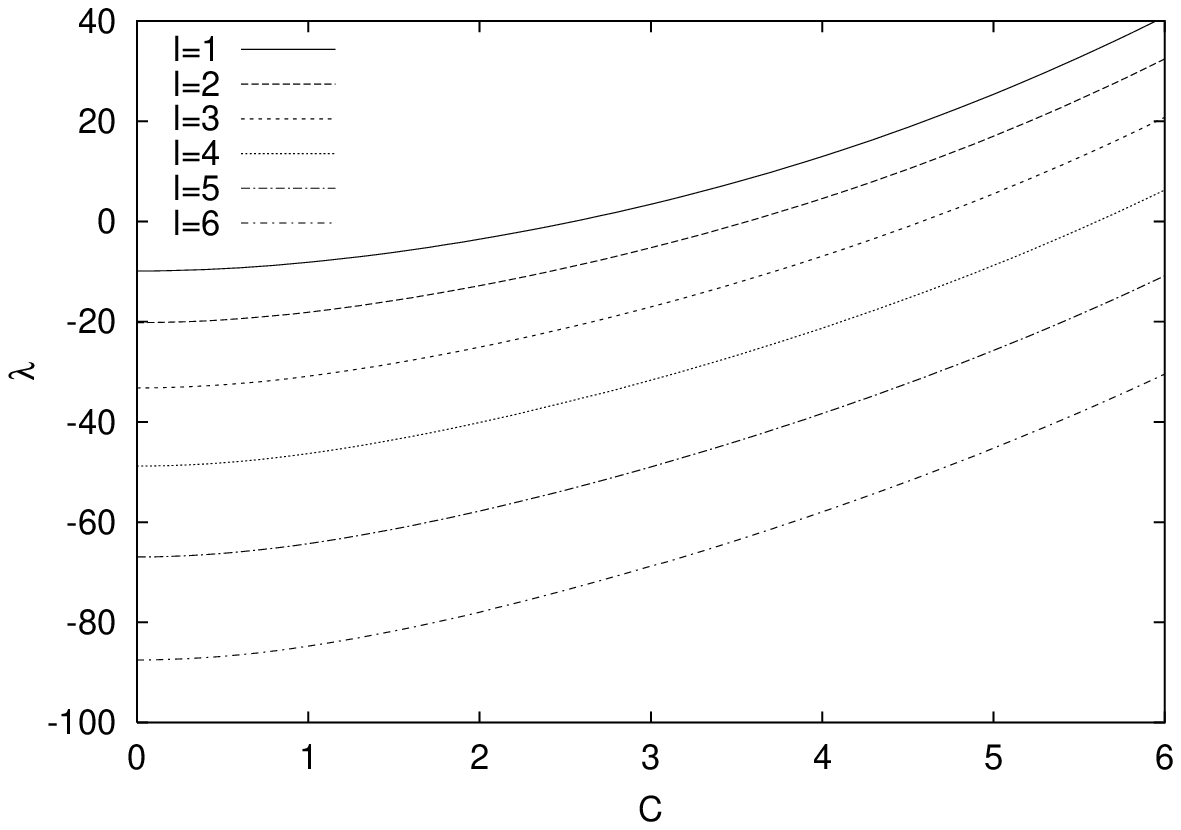} 
&\hspace{1.cm}\raisebox{35mm}{c2}&
\hspace{-1.cm} \epsfxsize=6cm\epsfbox{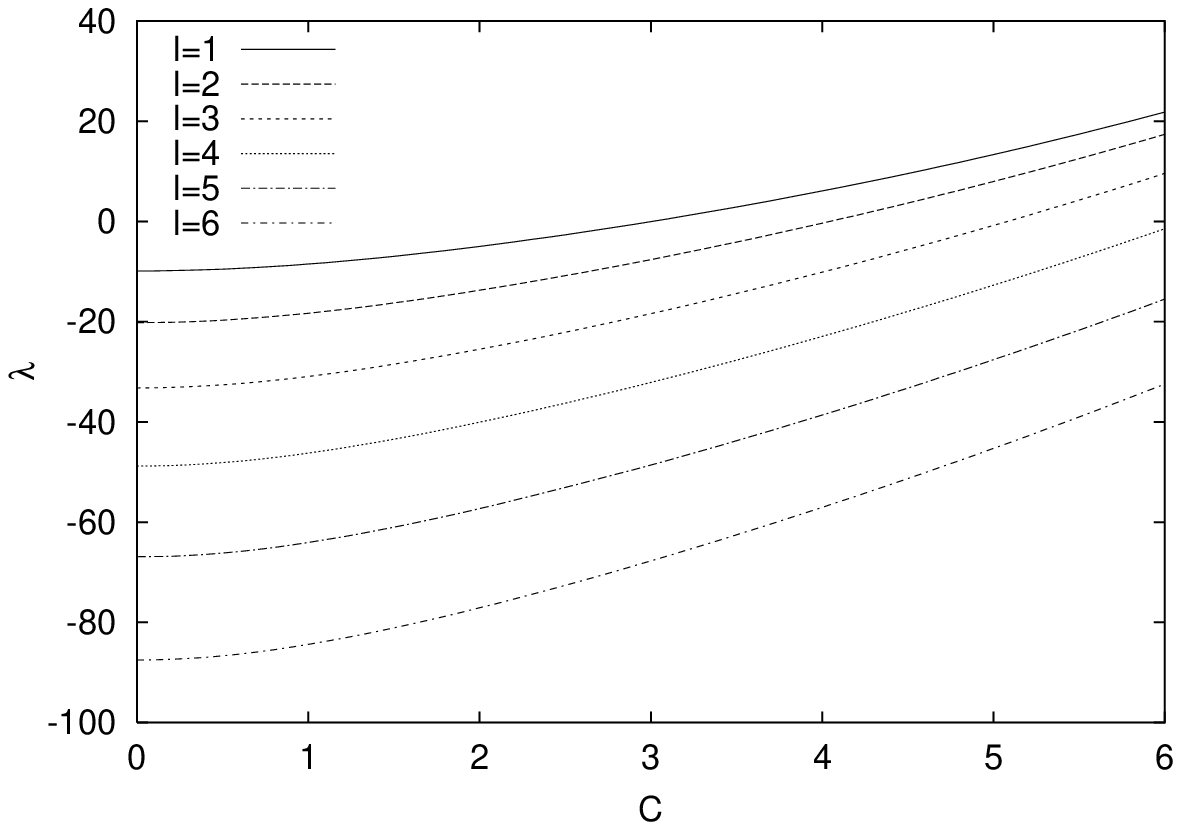}\\
\raisebox{35mm}{d1}&
\hspace{-1.cm}\epsfxsize=6cm\epsfbox{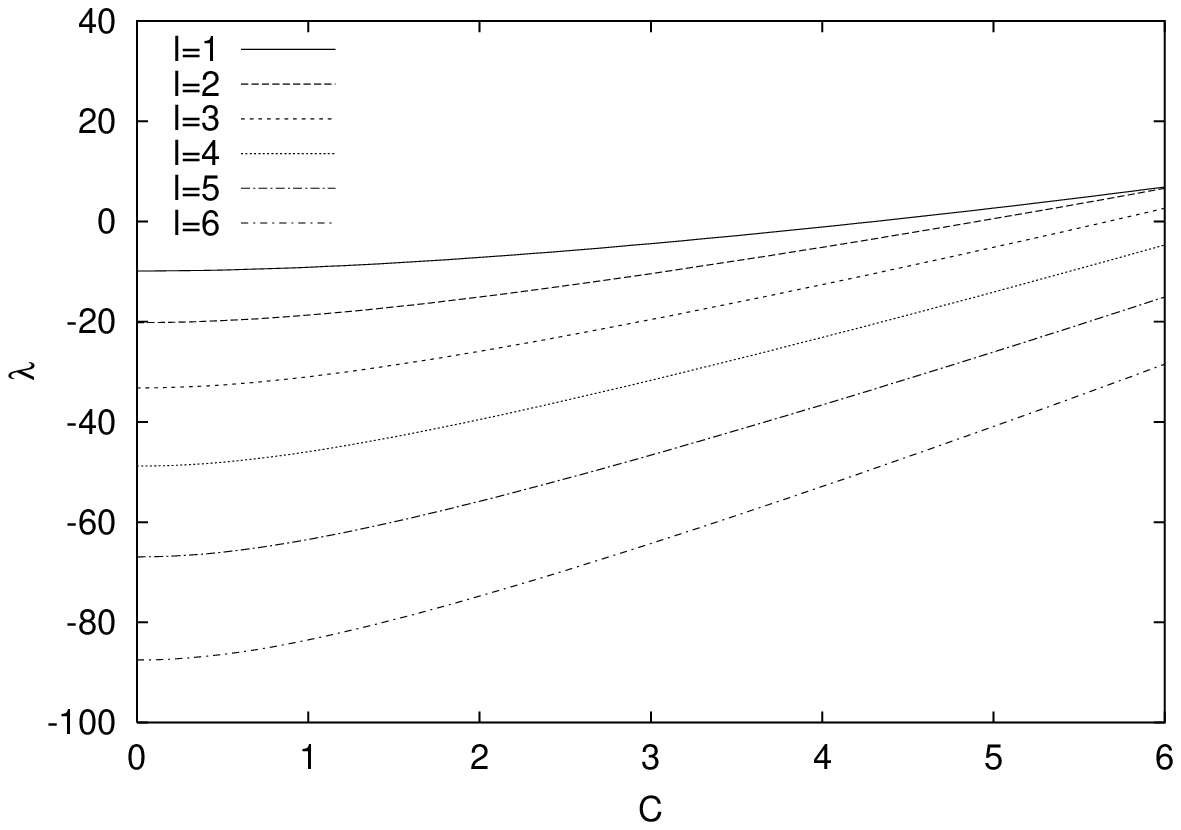} 
&\hspace{1.cm}\raisebox{35mm}{d2}&
\hspace{-1.cm} \epsfxsize=6cm\epsfbox{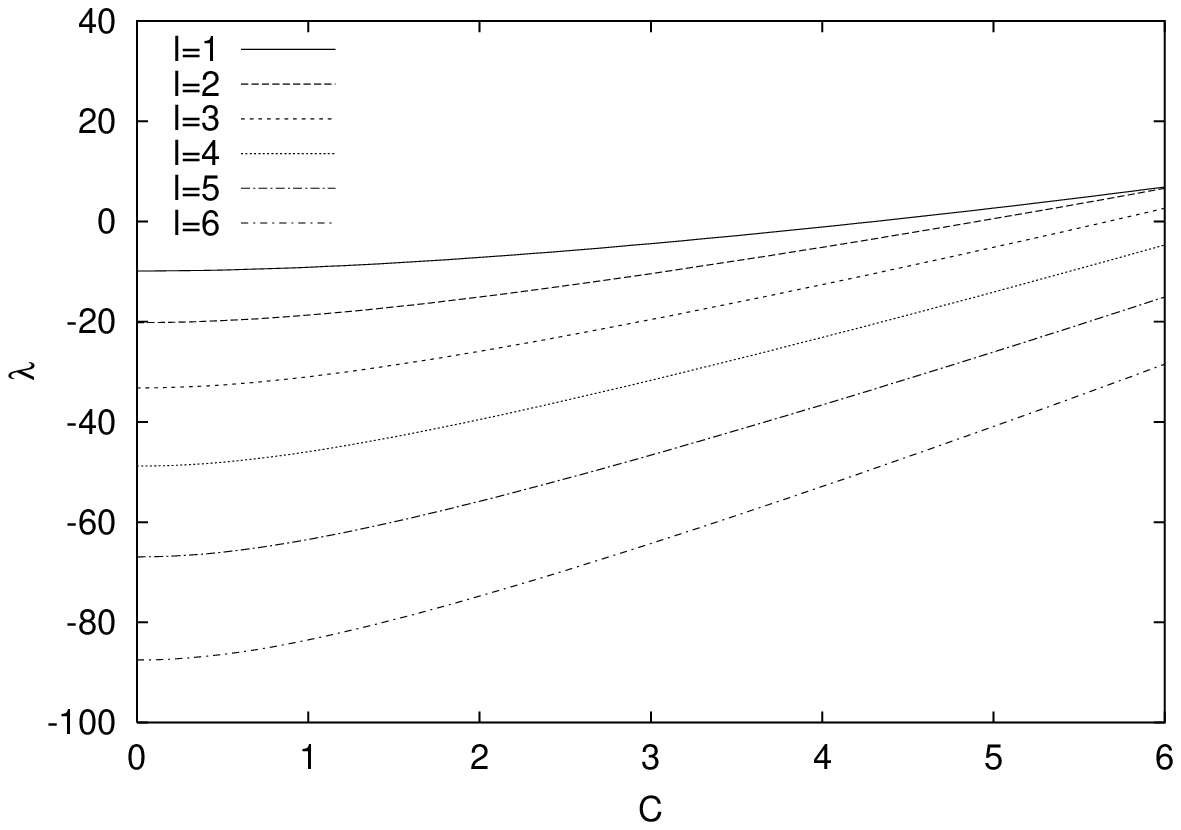}\\
\raisebox{35mm}{e1}&
\hspace{-1.cm}\epsfxsize=6cm\epsfbox{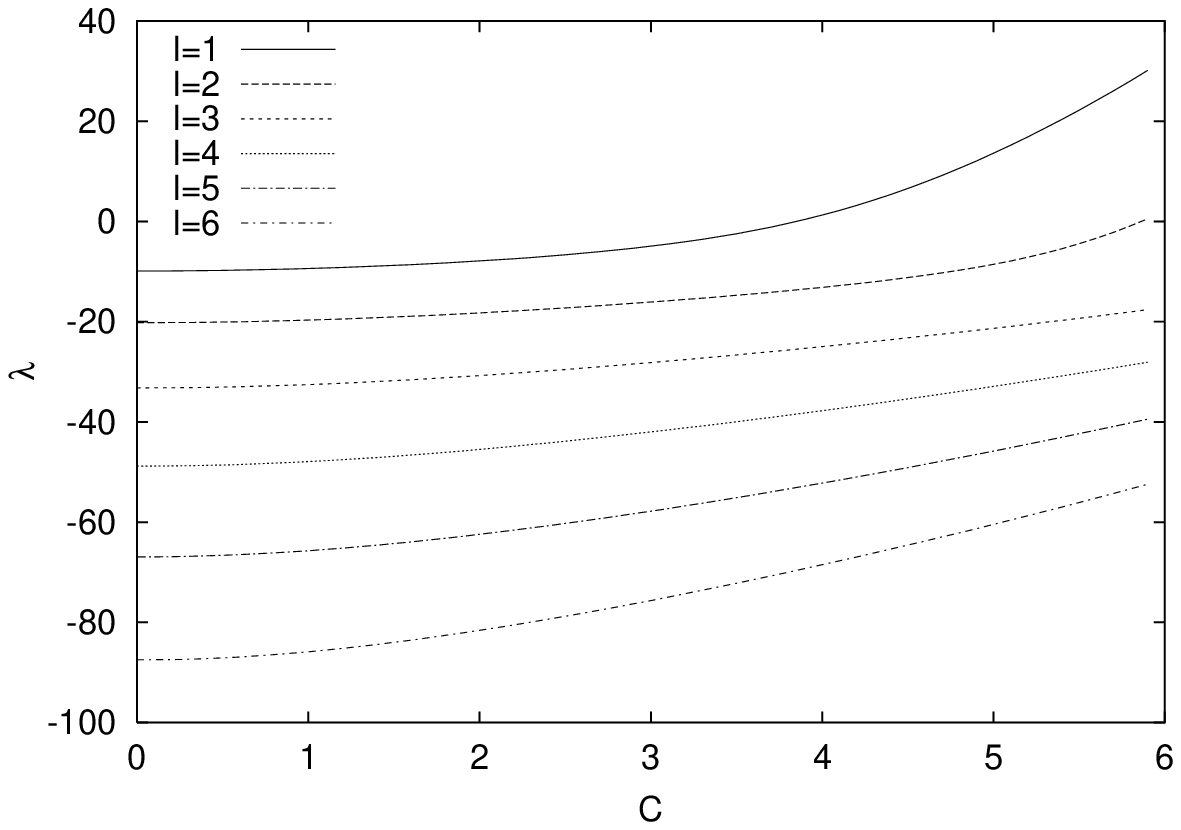} 
&\hspace{1.cm}\raisebox{35mm}{e2}&
\hspace{-1.cm} \epsfxsize=6cm\epsfbox{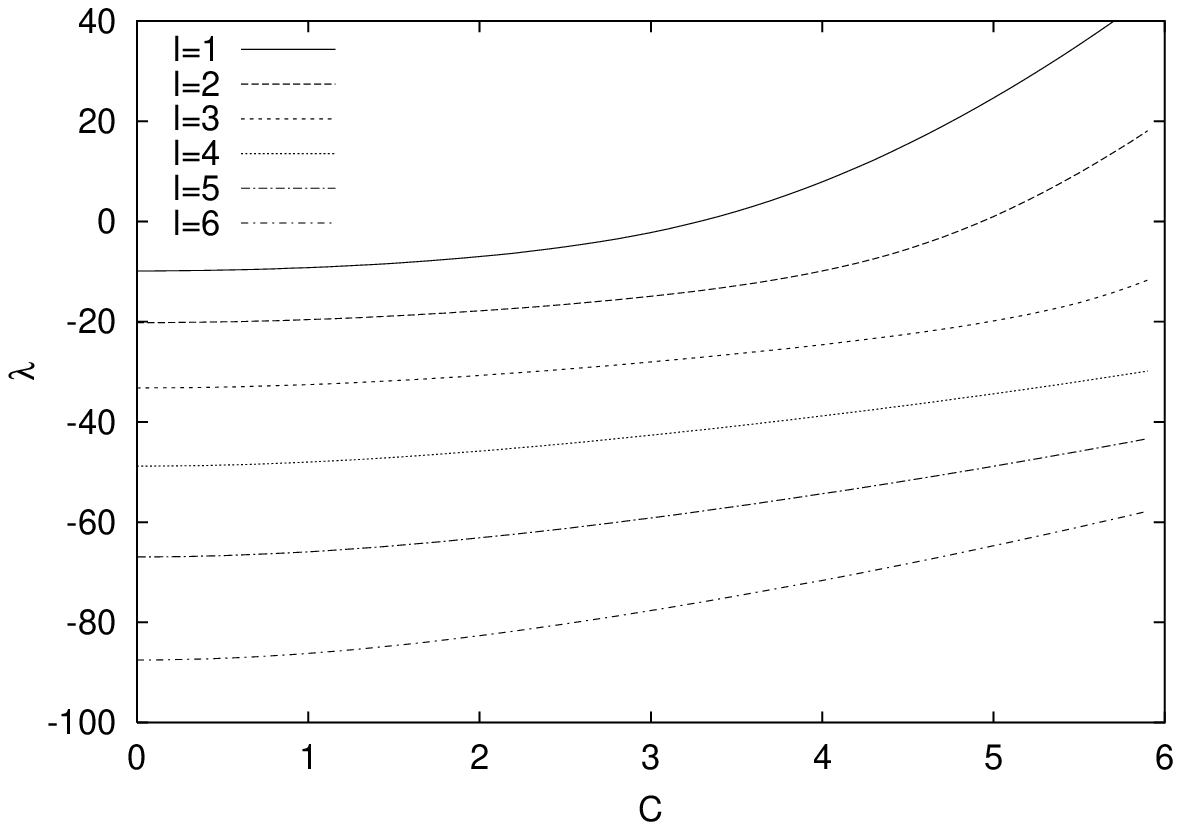}
\end{tabular}
\end{center}
\caption{Growth rates for $1 \le l \le 6$ for the  
models given in table 1.}
\end{figure}

Fig. 1 gives clear evidence of a dependence of 
the curves for different 
values of $l$
and their relations to each other on the function
$\alpha(r)$. For example, a monotonic increase 
of $\alpha$ (as in b1, b2) and a 
maximum between $0$ and $1$ (d1, d2) seems to
make the curves  for different $l$ ''converging'' 
to each other. On the other
hand,  monotonic decrease (c1, c2) or a 
minimum (e1, e2) seems to make the
curves more parallel or even more moving away. 
However this might be (and we
will see later that this behaviour is not universal) we 
only note here that there are clear differences in the 
relations of the eigenvalues of different field modes 
which will be
used in an inversion scheme to determine $\alpha(r)$.
Apart from example a2, we used only curves with at most 
one maximum or minimum.
In the next section we will face the problem 
that some oscillations 
(with at least one maximum and one minimum) 
around the treated 
paradigmatic model give eigenvalues which 
are (at least) very close 
to the
original one.

Note that the restriction to values 
$C<6$ was not voluntary as we have found 
(for some of the models) 
regions of $C$ were
no $\it{real}$ eigenvalues exist. This gives evidence 
for the
existence of oscillatory dynamos. 
This possibility
has been discussed in a number of papers (R\"adler 1986, 
R\"adler and Br\"auer 1987)
and is 
of some interest in its own right. Only for the sake of 
numerical 
simplicity we skip this problem here. There are no 
principle obstacles
to make the eigenvalue code fit for complex computations, 
too. As for the
inverse problem, which will be treated in the next 
section, we only 
remark that in case that some intermediate trial 
function of $\alpha(r)$
are found 
to have no real eigenvalues, these configurations are skipped and
are replaced by the next trial function.

\section{The inverse problem}

\subsection{Evolutionary Strategies}

In this section, we want to present and apply an inversion scheme
using  ''measured'' growth rates for a finite number of 
magnetic  field modes to get the radial 
dependence of $\alpha$. From the very beginning,
we will take into account the possibility that there could exist
various
solutions of this kind of inverse problem, 
i.e. various $\alpha$-profiles which can be 
distinguished considering, e.g., their 
mean quadratic 
curvature.

For that reason,  we decided to use a 
regularization of the inverse problem. One convenient  
way of regularizing
an inverse problem is to add to the usual functional of the 
mean squared 
residual deviation an additional weighted functional of  some 
appropriate
norm of quantities one is looking for and to search for minima
of the arising total functional. Using this so-called 
Tikhonov regularization
approach (for an overview see, e.g., Hansen 1992), 
one can scale through the regularization parameter. 
This way one usually gets  Tikhonov's L-curve. 
At the point of 
highest bending (the ''knee'') one can find a 
reasonable compromise between an optimal 
fitting of the model data to the measured data 
and a minimal norm of the
desired quantity. 
 
Here, we will slightly modify this regularization method
in the following way. To begin with, we restrict 
the space of functions to polynomials of 
fourth order, excluding the
term proportional to $r$ ($\alpha(r) \sim r$ would correspond 
to a cusp of $\alpha$ at the origin which seems to be
not very realistic). 
These polynomials are parameterized  by  values  $\tilde{\alpha}_i$
at the fitting points 
$r_i=0.25,0.5,0.75,1.0$ where the polynomial expansion of $\alpha(r)$ 
has to fit
optimally the values $\tilde{\alpha}_i$. 
In this first fitting procedure, however, we use 
an additional 
weighted functional which keeps the mean squared 
curvature 
more or less small. For a high regularization parameter 
any set of (four) values
at the  fitting points leads to relatively smooth curves in 
the space of polynomial expansion 
coefficients. When the regularization parameter goes to 
zero the fitting
curve tends exactly to the parameterizing values at the 
fitting points. 
Thus, we are able to regularize the curves even before they are 
put into the kernel of the inversion scheme.

To put this procedure in formulae, let us assume that we have 
a trial set of values $\tilde{\alpha}_i$ at 
the fitting points $r_1=0.25$, $r_2=0.5$, $r_3=0.75$,
and $r_4=1.0$. Fitting a function 
$\alpha(r)=a+b r^2+c r^3+dr^4$ will be
done by minimizing the functional
\begin{eqnarray}
F=\sum_{i=1}^4 (\alpha(r_i)-\tilde{\alpha}_i)^2 +p 
\int_0^1 (2 b+6 c r
+12 d r^2)^2 dr
\end{eqnarray}
where $p$ denotes the regularization parameter.

After having explained the chosen parameterization, 
we will describe in 
the following the chosen minimum search  procedure. 
We decided to use an evolutionary strategy (ES) for that purpose 
(for an overview see, e.g., Sch\"oneburg, Heinzmann and 
Feddersen 1994). An 
ES is 
known to be an appropriate method in case of existence 
of several
local minima. It is capable to leave local minima in 
order to find
the global minimum of a given problem. We will 
sketch the philosophy
of the used ES only shortly. For our particular code, 
we took over the scheme described by K\"uchler, Lehnert and 
Tschornack (1995).  

An ES tries to utilize the principle of 
biological evolution  for the numerical
solution of (non-linear) optimization problems. Let us 
start with an 
''population'' of 30 ''individuals'' which are 
vectors of parameterizing values at the fitting
points. To each of those individuals a quality function $Q$ is 
ascribed. In order to compute this quality function $Q$, we 
have to solve, for
every considered $l$, 
the eigenvalue equation for the trial function 
$\alpha(r)$ which results from the 
vectors of parameterizing values the way described above.
Having computed the eigenvalues for the given individual
we compare these eigenvalues with the measured ones and define the
quality function as the mean squared residual deviation. 
Here, we assume the a-priori 
errors for all $l$-mode measurements to be equal. Of course, 
this  
might be corrected in
any practical application. 

The population 
evolves in the following manner: In every 
new generation only one
child is created by one of three different kinds of 
reproduction which are
mutation, cross-breeding, and crossing-over. For every
generation the actual reproduction kind is 
used by random. For the relative 
frequency of mutation, cross-breeding, and crossing-over we 
have chosen
the values 0.3, 0.3, and 0.4, respectively.

Mutation means that we select one ''parent'' by 
random and create from this
a ''child'' by mutation of the parameters, i.e.
\begin{eqnarray}
\tilde{\alpha}_i^C=\tilde{\alpha}_i^P+\Delta_i \gamma s \; .
\end{eqnarray}
Here $\tilde{\alpha}_i^C$ is the i'th parameter of the child, 
$\tilde{\alpha}_i^P$ is the i'th parameter of the 
parent, $\Delta_i$ is the
spread of the corresponding parameter, $\gamma$ is
a mutation width and $s$ is a normal distributed random number. 
It should be noted that $\Delta_i$ is large at the beginning 
of the 
evolution when a large area of the parameter space is 
to be covered, but that
it decreases when the whole population is running into 
the global minimum.

Cross-breeding means that  the parameters of two 
parents $P1$ and $P2$ which are selected by random are 
averaged in the child:
\begin{eqnarray}
\tilde{\alpha}_i^C=0.5\; (\tilde{\alpha}_i^{P1}+
\tilde{\alpha}_i^{P2}) \;\;\; .
\end{eqnarray}

Two parents are also selected in the crossing-over, 
but their parameters
will be mixed in the sense that for each parameter 
$\tilde{\alpha}_i^C$ it will be  selected by random 
whether it comes 
from parent $P1$ and $P2$.

After having created a child, it must be evaluated. 
At first, it has to be
checked whether the parameters of the child still 
fulfill some 
reasonable 
given constraints. If this is not the case, a new 
child has to be created. 
In the positive case, however, it is checked whether 
the quality function of the
child fulfills the condition
\begin{eqnarray}
Q_c<Q_{worst}+0.4 \; (Q_{worst}-Q_{best})
\end{eqnarray}
where $Q_{worst}$ and $Q_{best}$ are the quality functions
of the worst and the best individual of the
population, respectively.
The value 0.4 is to some extend arbitrary. Evidently this factor 
is responsible for the fact that the ES is able to leave 
local minima. 
The chosen factor 0.4 is relatively high. The effect is  
a certain slow-down 
of the convergence but a higher reliability to find the 
global minimum 
and not to be locked in a local one. 
If the condition (13) is fulfilled, after cross-breeding 
and crossing-over the 
qualitatively worse parent is replaced by
the child. In case of 
mutation, the parent is replaced, except when the parent is the 
best individual of the population and the quality of the child is 
not better than the quality of the parent. In this special case, 
the worst individual of
the population is replaced by the child.

\begin{figure}[t]
\begin{center}
\begin{tabular}{ccc}
\raisebox{35mm}{a1}&\epsfxsize=6cm\epsfbox{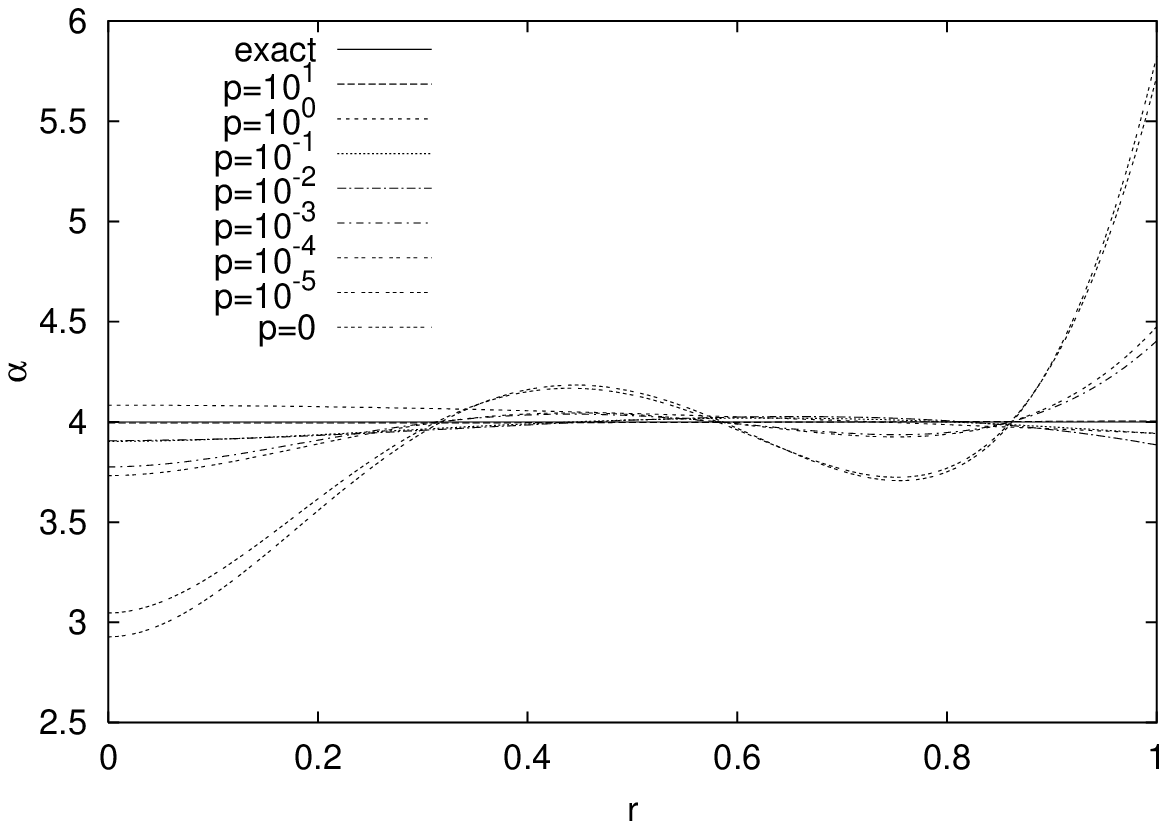} & 
\epsfxsize=6cm\epsfbox{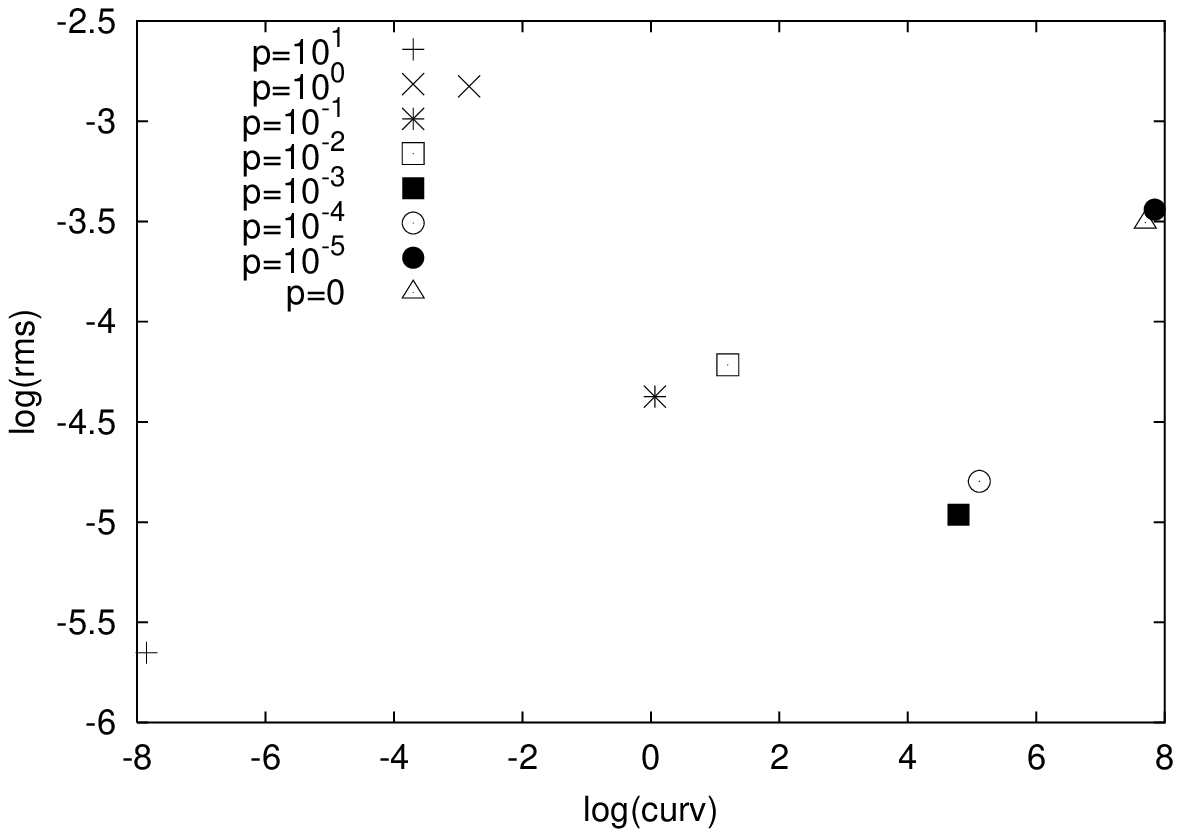}\\
\raisebox{35mm}{b1}&\epsfxsize=6cm\epsfbox{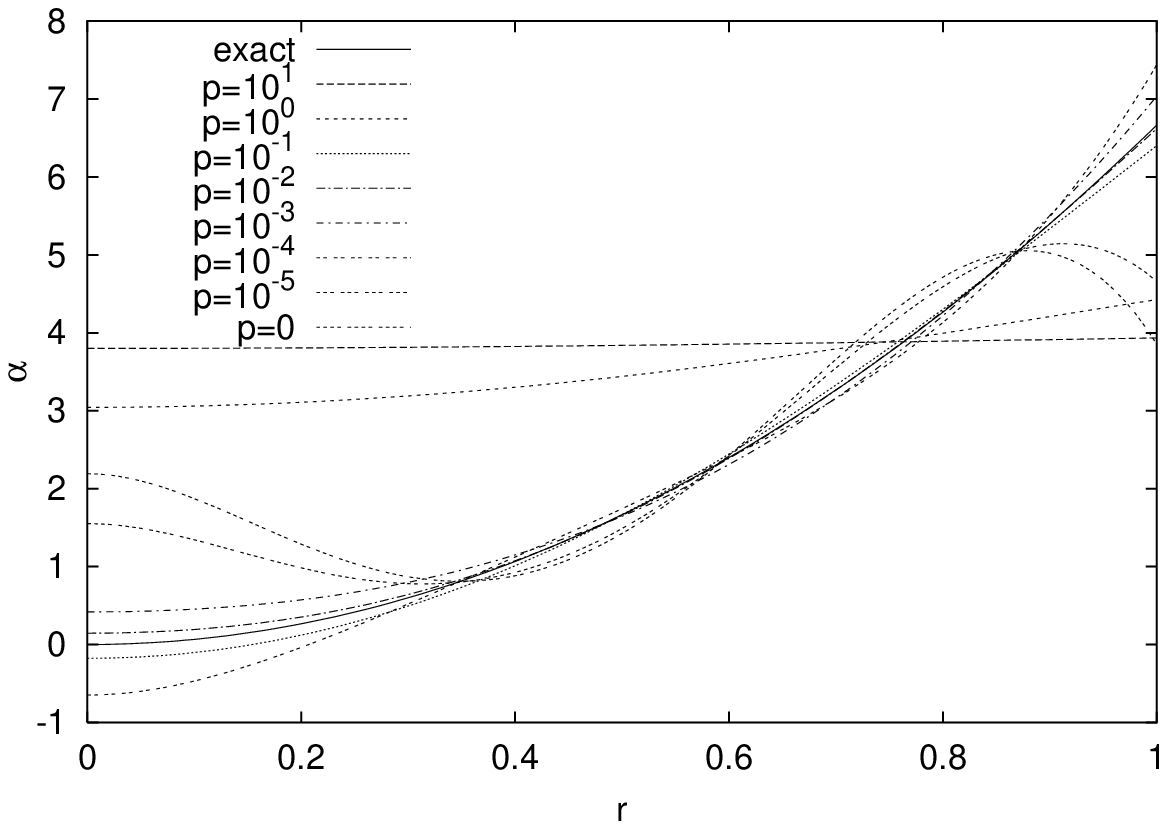} & 
\epsfxsize=6cm\epsfbox{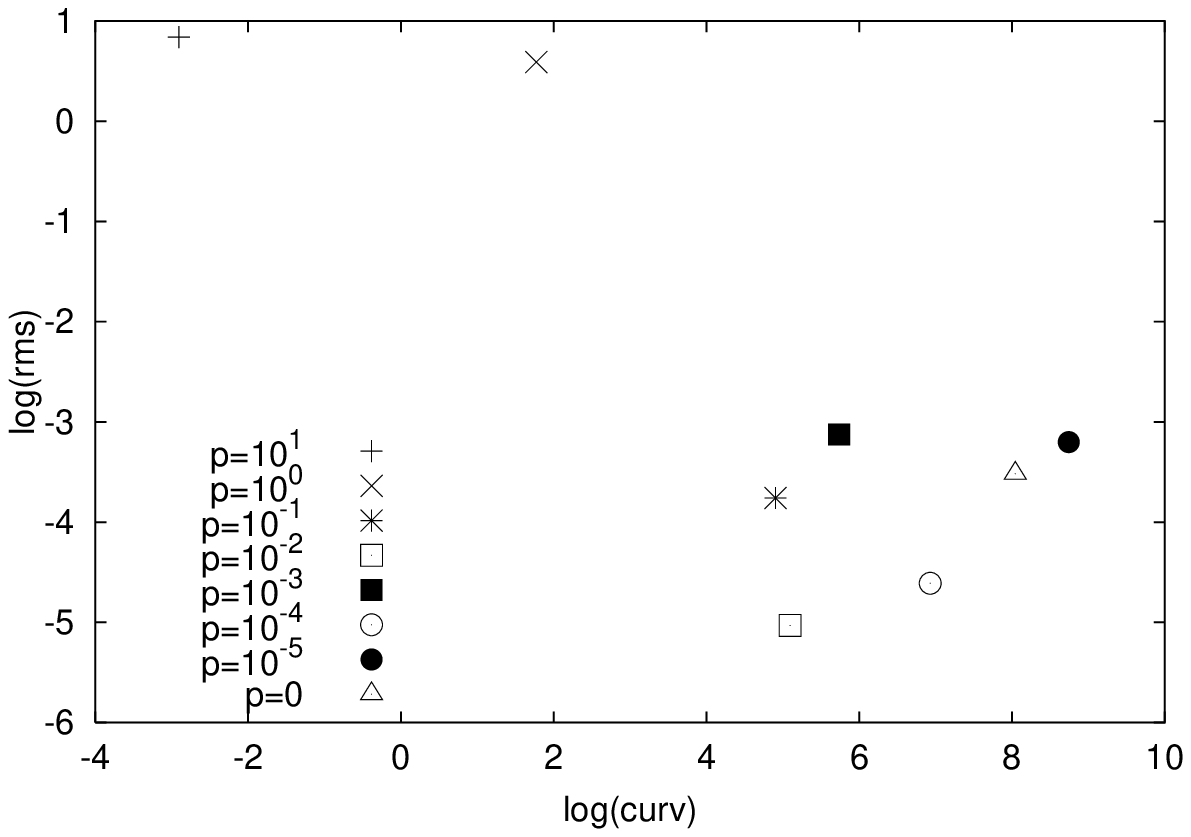}\\
\raisebox{35mm}{c1}&\epsfxsize=6cm\epsfbox{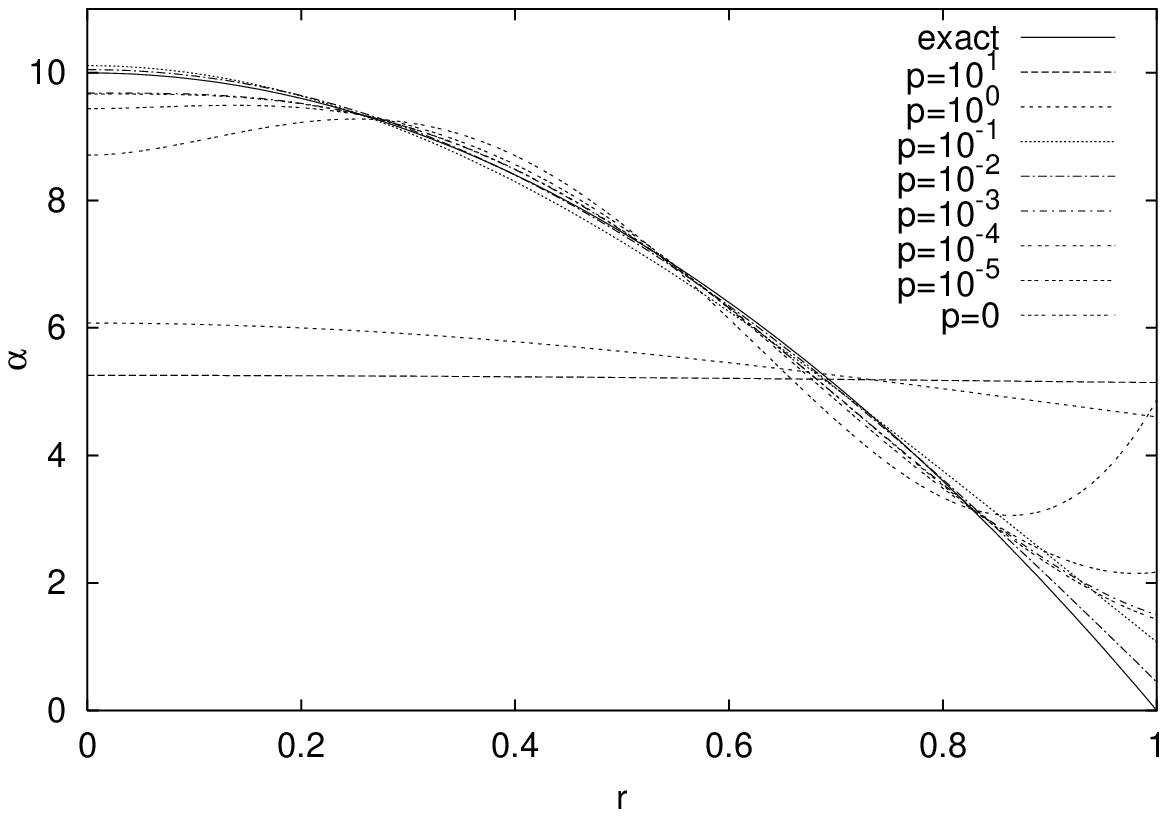} & 
\epsfxsize=6cm\epsfbox{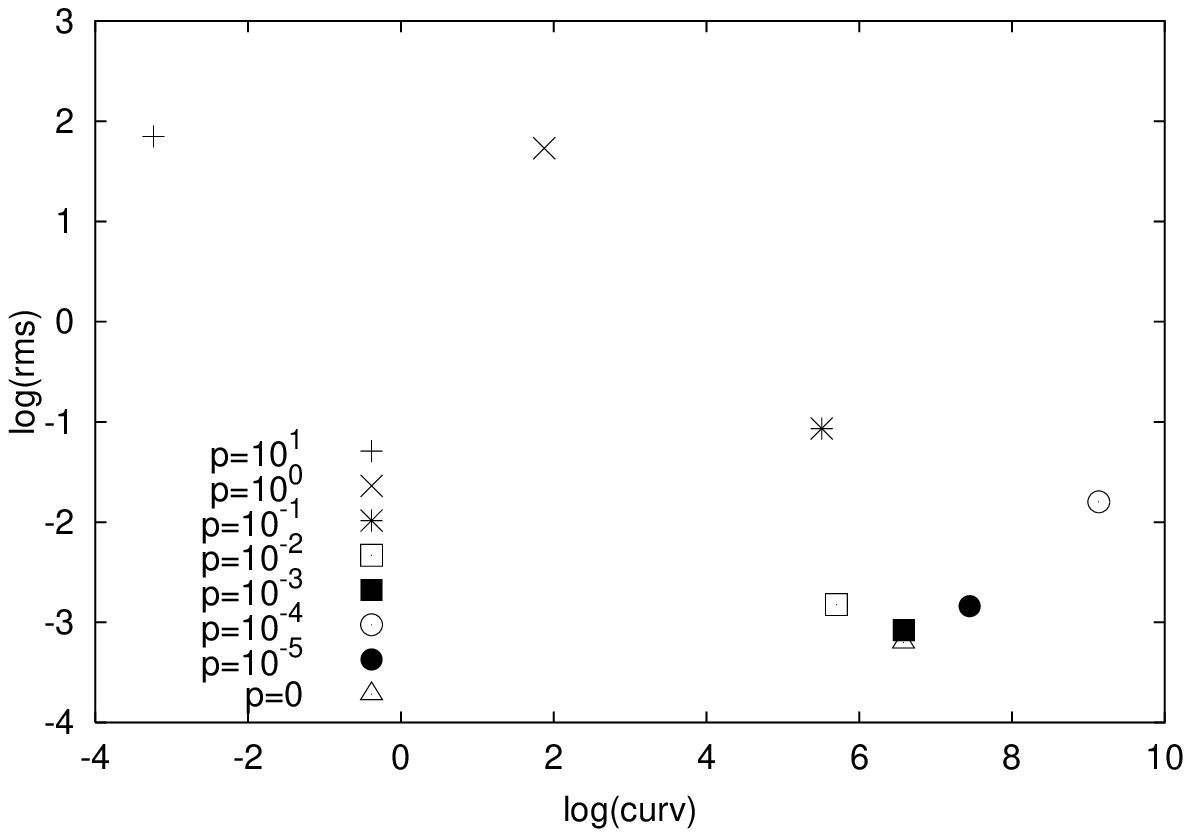}\\
\raisebox{35mm}{d1}&\epsfxsize=6cm\epsfbox{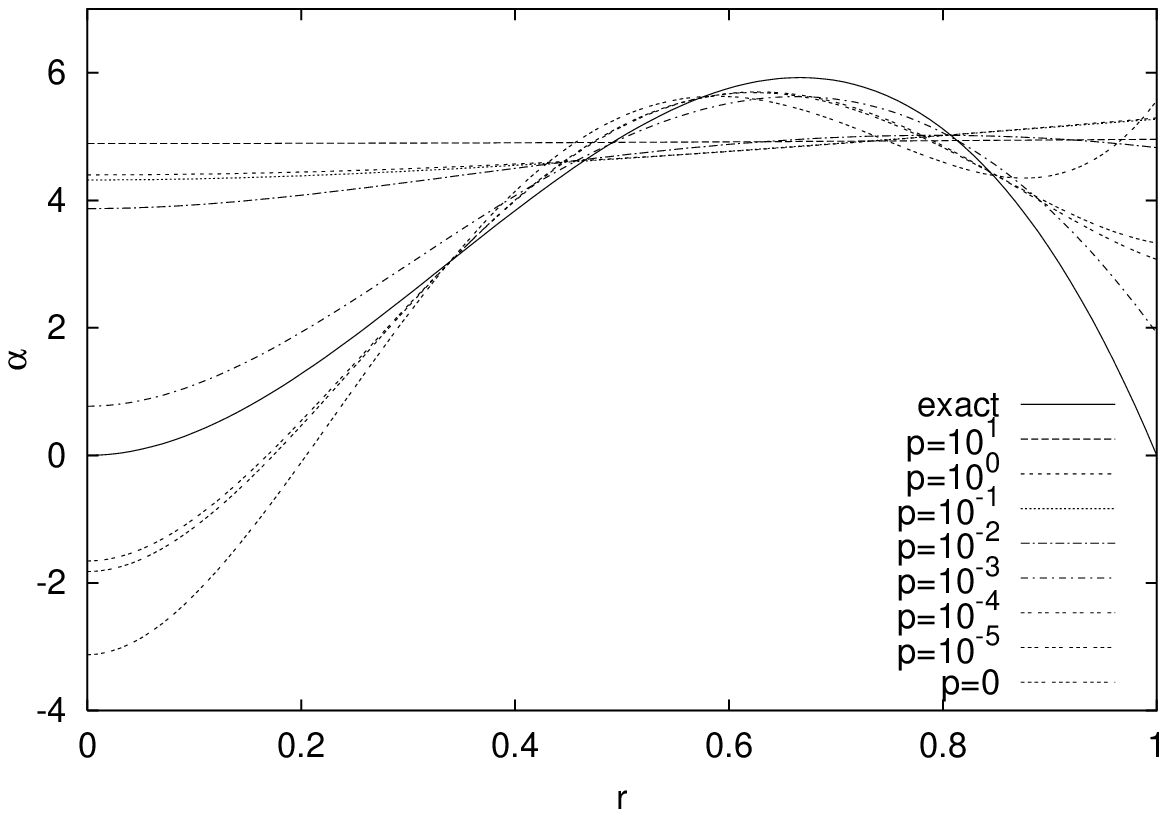} & 
\epsfxsize=6cm\epsfbox{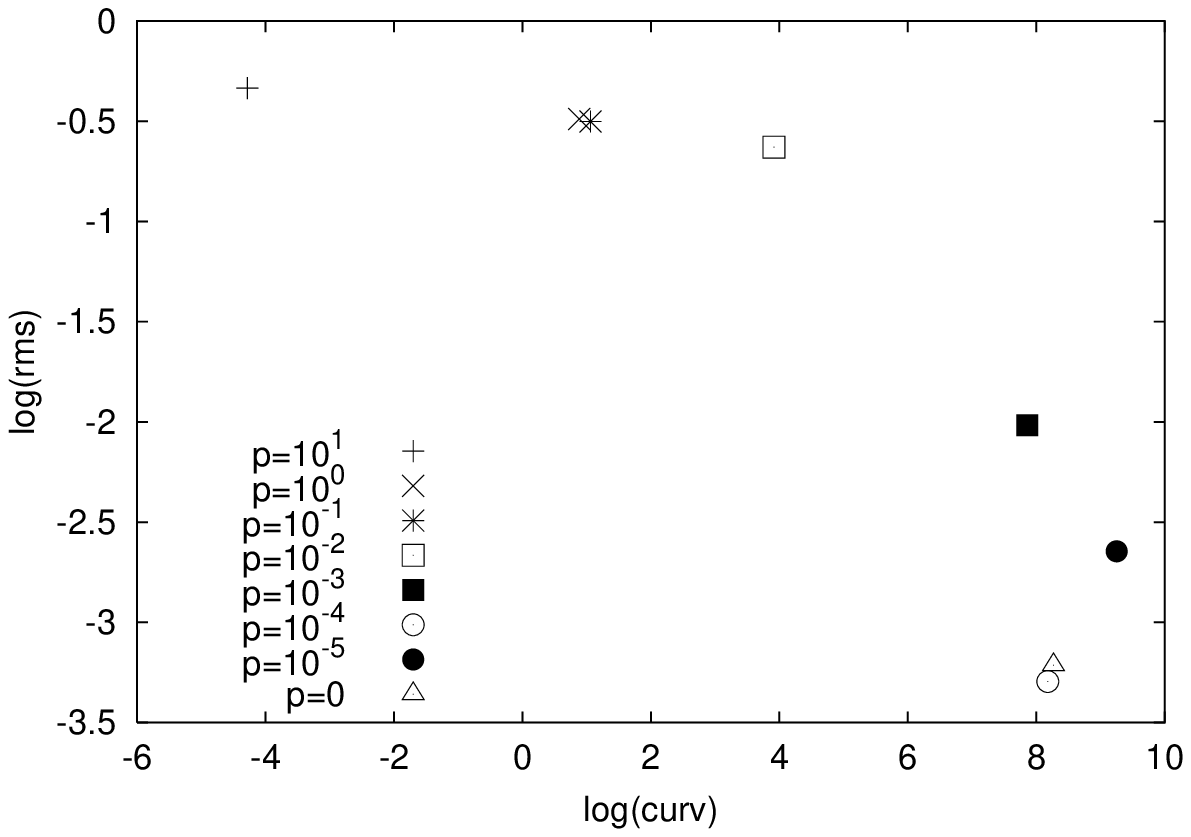}\\
\raisebox{35mm}{e1}&\epsfxsize=6cm\epsfbox{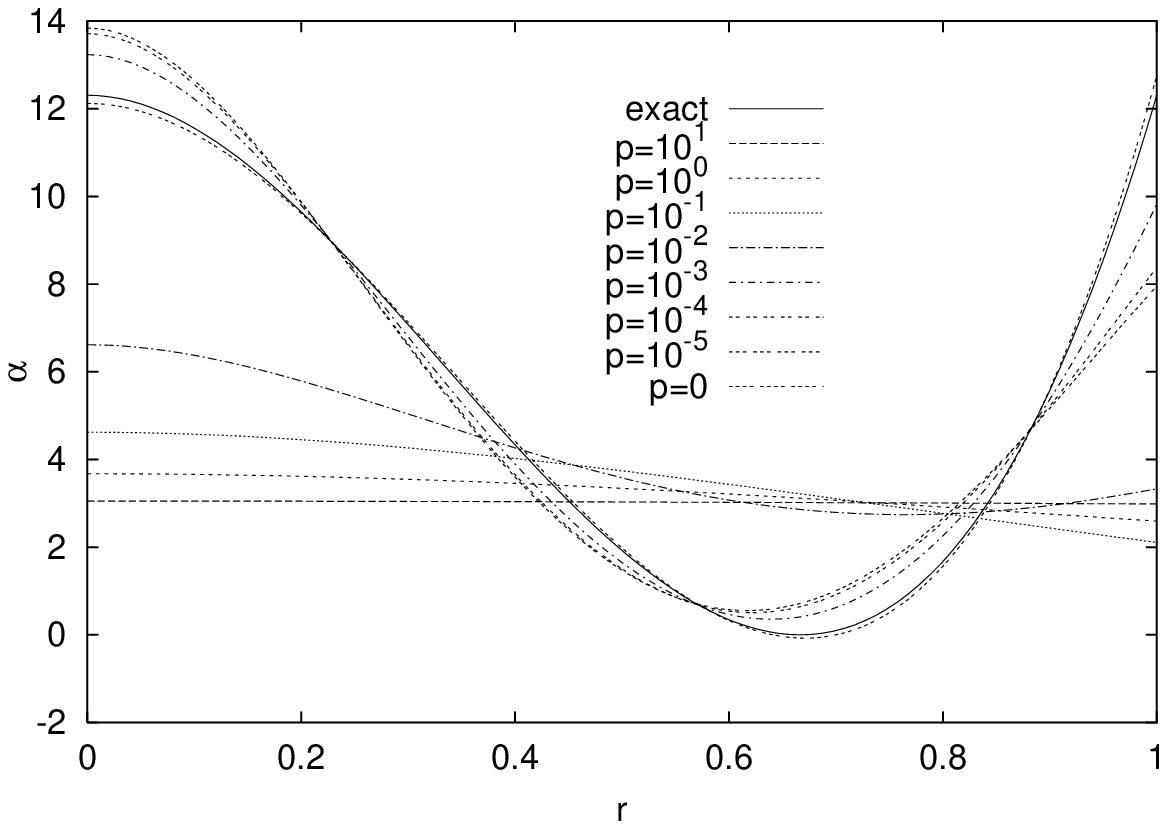} & 
\epsfxsize=6cm\epsfbox{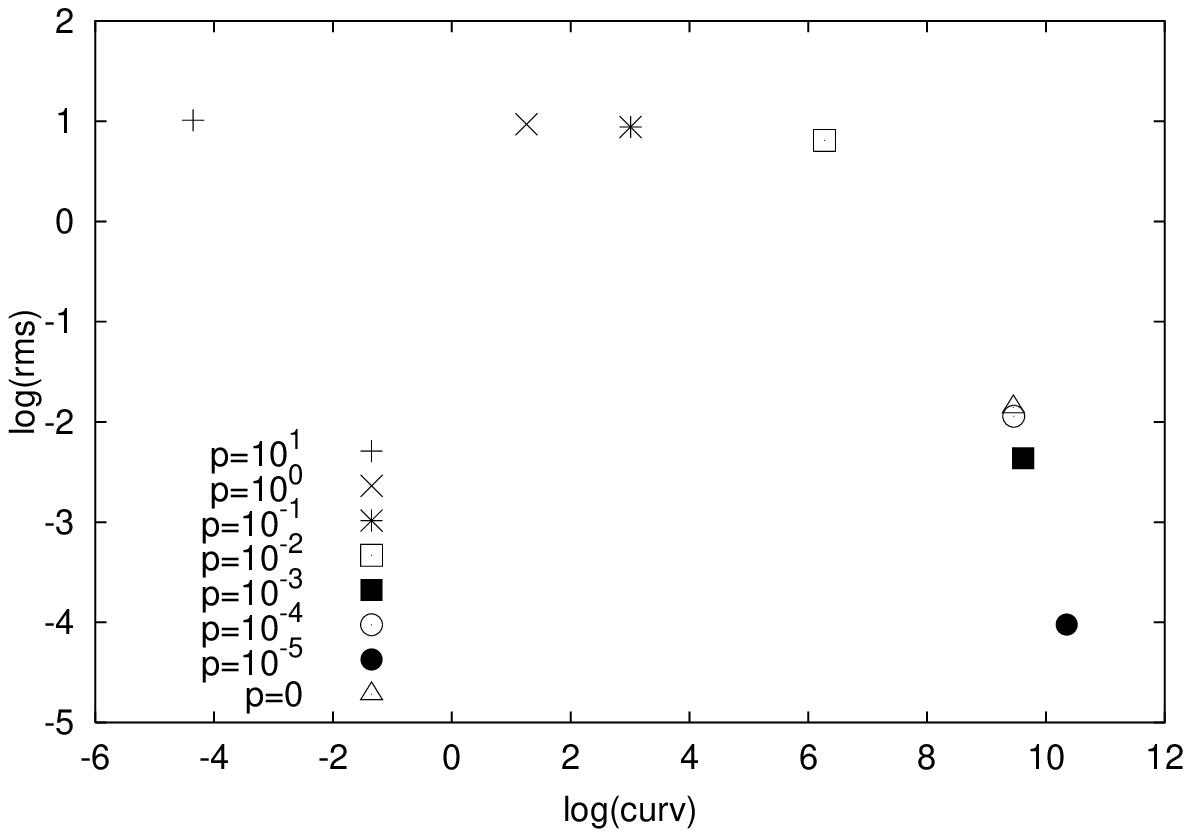}
\end{tabular}
\end{center}
\caption{Left column: original functions $\alpha(r)$ and 
functions resulting from the ES inversion procedure for 
different regularization 
parameters  $p$.
Right column: dependence of the rms of the residual 
error on the
mean squared curvature of the resulting function.
The cases a1, b1, c1, d1, and e1 correspond to the models 
considered in section 2.}
\end{figure} 

Evidently,  the number of individuals in every generation stays 
the same. 
The evolution can be stopped when all individuals of the population
have gathered in one (hopefully: the global) minimum, e.g., 
when the spread of all
parameters has become very small. Of course, if several 
minima of the 
quality function with nearly the 
same depth exist, the evolution might be locked in 
quite different of
them, depending
on the random numbers used in the evolution. In order to 
control this 
erratic behaviour, we have used the regularization 
scheme described 
above. Thus, scaling the regularization parameter $p$ from 
larger to 
smaller values, we expect  the smoothest solution to 
show up first and
solutions with higher curvature to appear only 
for smaller values 
of $p$. Than, however, the mentioned erratic behaviour 
might be observed.

\subsection{Results for the paradigmatic examples}

Fig. 2 shows the results of the used ES inversion 
scheme for the examples
a1, b1, c1, d1, and e1 from section 2. Every plot 
on the right hand side 
shows nine curves $\alpha(r)$, the first representing 
the exact input function
which was used in the forward task for the determination of the
growth rates for different $l$-modes. The remaining eight curves 
give the output of the ES inversion scheme, 
differing in the value of the used regularization
parameter $p$ in equation (9). As input for the inverse task, we
have used the growth rates for 
the first 6 $l$-modes at $C=4$. Reminding
that we parameterize
the desired functions $\alpha(r)$ by 
means of 4 values, 6 growth rates 
give evidently some redundant 
information.

The plots on the right hand side of 
Fig. 2 are similar to what is
called in regularization theory 
''Tichonov's L-curve''. On the abscissa
the logarithm of the squared curvature of 
the output curves is given. 
The ordinate axis shows the logarithm of the 
rms of the residual deviation 
(i.e., the difference of the ''measured'' growth rates and the
growth rates resulting from  the output curves). For further 
explanation, let us 
consider example b1 
having a quadratic $\alpha(r)$-dependence as original function 
(example a1 is atypical, as the original 
curve has curvature zero). To begin with, consider in b1 
the curves  for
$p=10$ and $p=1$. Evidently, the resulting functions $\alpha(r)$
are close to a constant, e.g., the 
large regularization parameter $p$ allows 
only for output curves of very small mean curvature. On the right 
hand side we see that the 
corresponding residual deviation for those 
curves are very large. Then, with decreasing $p$, 
we get three curves 
which are fitting reasonably to the original one. The corresponding 
residual deviations are decreasing drastically. 
But for the last three
values of $p$ ($p=10^{-4}$, $p=10^{-5}$, and $p=0$) the resulting 
functions $\alpha(r)$
acquire some oscillatory behaviour. Looking at the right hand side 
plot, we see that the residual deviations for these functions are a 
bit higher compared 
to the minimum, but evidently the ES is not able to 
find the global 
minimum. 
A quite similar behaviour is found for all other examples (note 
that for example a1 ''Tikhonov's L-curve'' starts already at 
high values of $p$
with a small residual error, 
as the original curve has no curvature).

\begin{figure}[t]
\begin{center}
\begin{tabular}{cc}
\epsfxsize=6cm\epsfbox{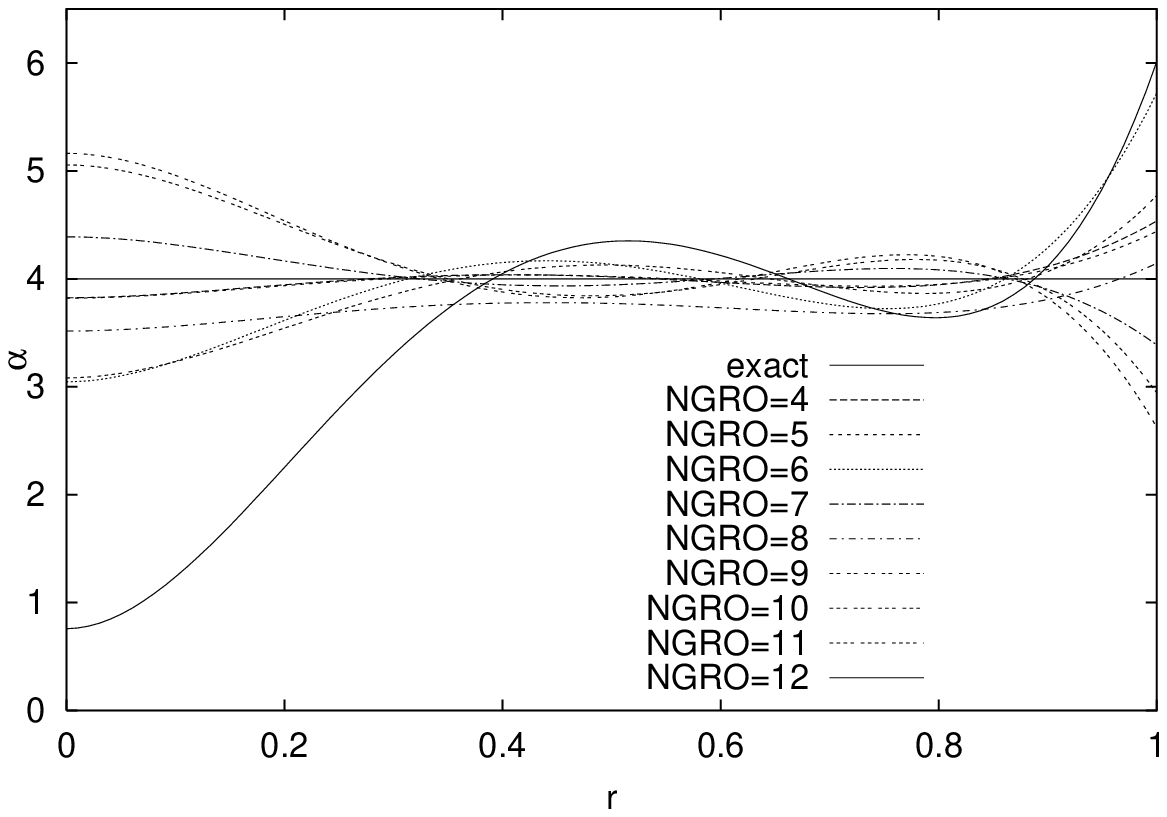 } & 
\epsfxsize=6cm\epsfbox{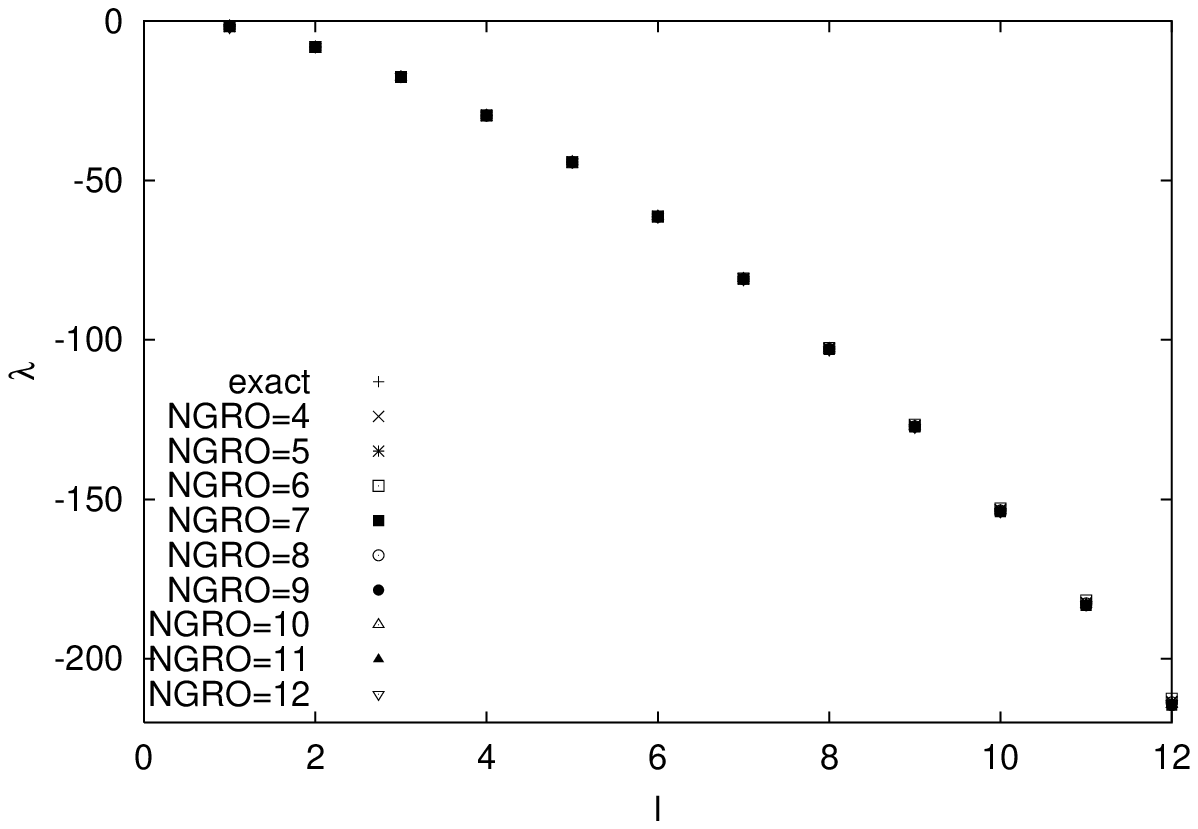}\\
(a)&(b)\\[4mm]
\epsfxsize=6cm\epsfbox{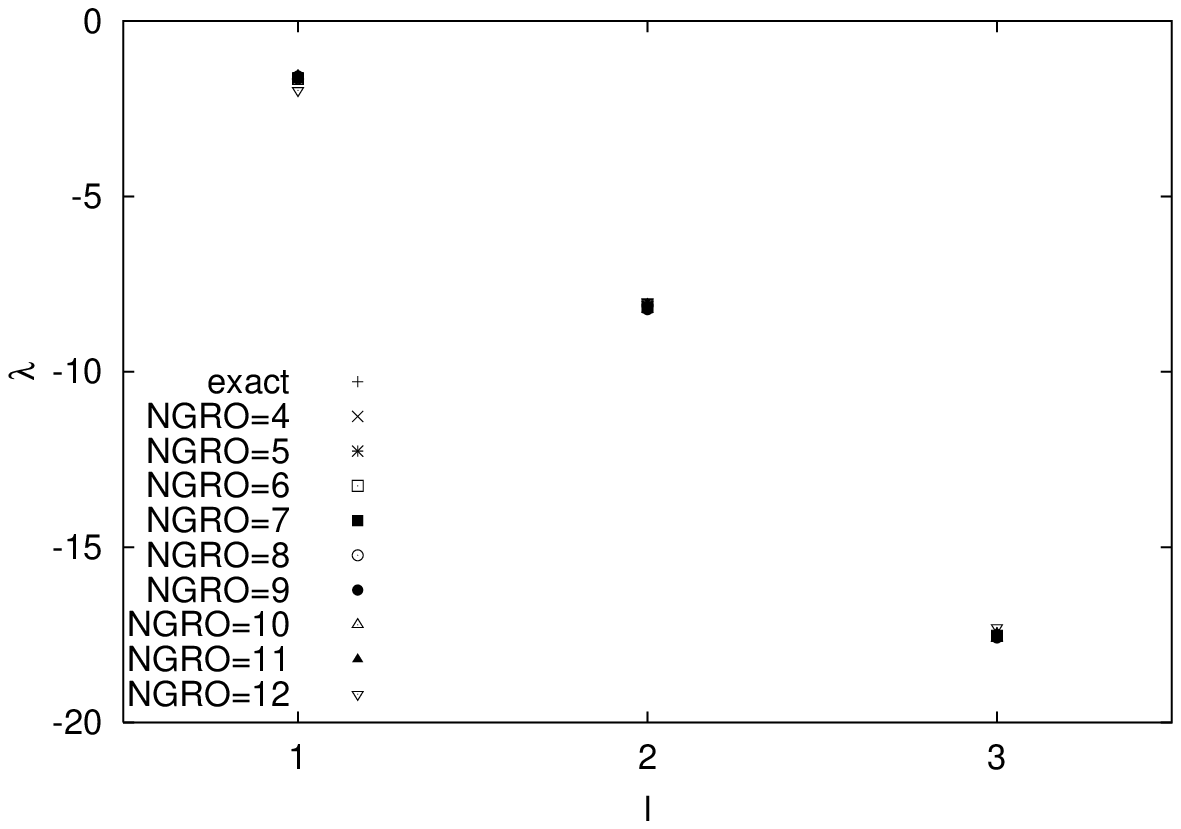 } & 
\epsfxsize=6cm\epsfbox{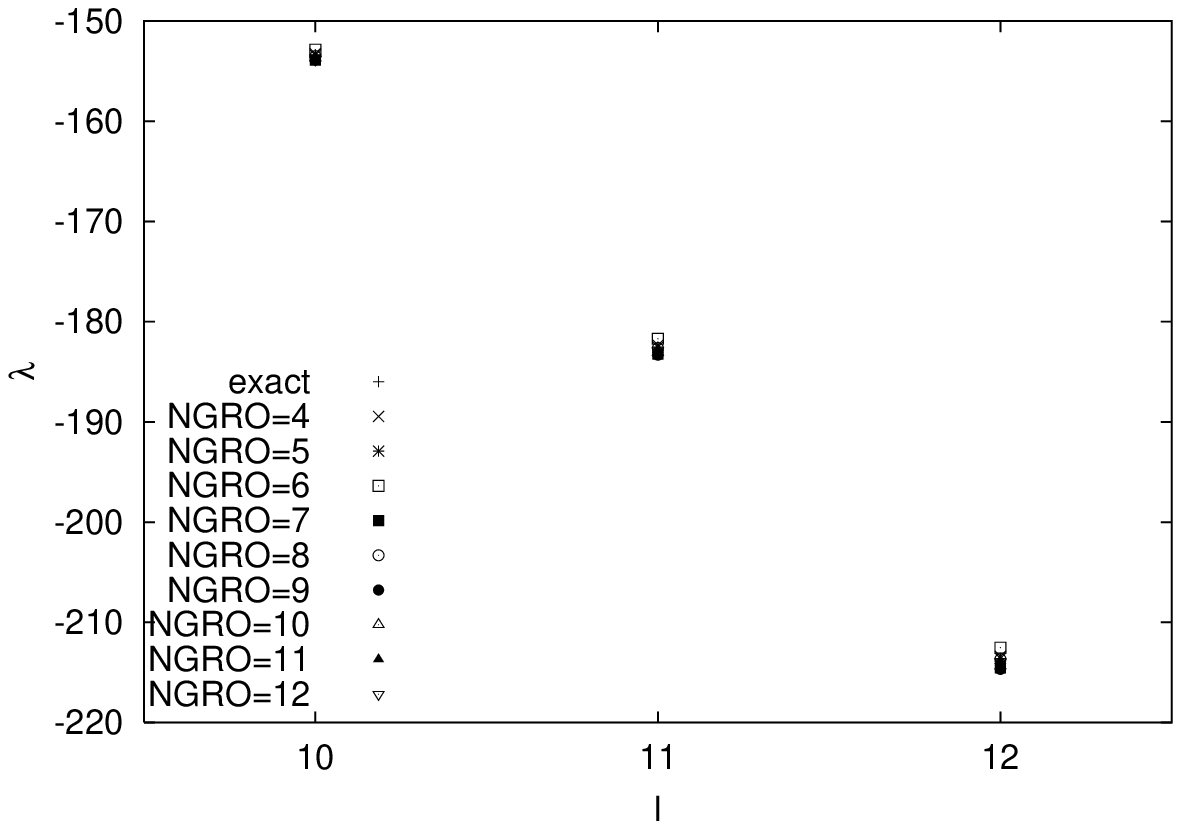}\\
(c)&(d)
\end{tabular}
\end{center}
\caption{Results of the ES inversion scheme for model a1 using 
different numbers $NGRO$ of ''measured'' growth rates for 
modes with $1 \le l \le NGRO$. The functions are shown in (a). 
In (b) the growth rates (for $1 \le l \le 12$ of all resulting
curves are plotted. The same in (c) and (d), but detailed 
for the lowest three $l$ and the highest three $l$. Evidently, 
the 
resulting growth rates for quite different functions $\alpha(r)$ 
are practically indistinguishable.} 
\end{figure}

One might argue that another minimum search code would be better 
suited to
find the absolute minimum.  Note, however, that every 
real measurement is biased by measurement errors. Considering the 
value of the rms of the 
residual deviations on the right hand side, it seems clear that
a realistic discrimination between the curves is hardly possible. 
Note, in addition, that 
our special parameterization (with a power series expansion up to
fourth order) might prevent curves for smaller 
regularization parameter  
from fitting 
to ''measured'' growth rates more exactly.

Connected with this problem 
one could also argue that the incorporation 
of more measured growth rates could improve the quality of the 
inversion. For this purpose, we carried out a numerical test taking
into account different total numbers (denoted by $NGRO$) 
of ''measured'' growth rates. 
For example a1 we have tested all possibilities 
with $4\le NGRO \le 12$ at $p=0$. Fig. 3a shows the resulting 
functions. 
Figs. 3b, 3c, and 3d give 
the corresponding growth rates for the
exact  function and all functions resulting from the inversion 
(Fig. 3b shows the values for all modes with $1\le l \le 12$, Fig. 3c shows 
the details for the first modes with $1\le l \le 3$, 
Fig. 3d shows the details 
for the modes with $10 \le l \le 12$). The result is astonishing:
all the quite different functions $\alpha(r)$ shown
in Fig. 3a give nearly the same growth rates for the 
first modes with $1\le l \le 12$, apart
from very small differences. Of course, our numerical method 
and the used parameterization 
with a very limited number of parameters 
does not allow to construct exactly those different functions 
$\alpha(r)$ which might give exactly the same eigenvalues. 
At this point we would like to suggest analytical 
investigations in order to identify these
isospectral functions $\alpha(r)$. 

For any practical application where the accuracy of
the measured values will be limited the
consequences are clear: there are quite different functions 
$\alpha(r)$ giving nearly the same growth rates. Thus, 
a unique determination of the radial dependence 
of $\alpha$ from  the eigenvalues of different $l$-modes
seems to be impossible. However, using some kind of regularization, 
we can find those $\alpha$-functions giving  the 
measured eigenvalues with minimal penalty function, e.g, with 
minimal quadratic curvature. Hence, it is  possible 
to reconstruct at least  $\alpha$-functions of the
paradigmatic types as described in section 2.

\subsection{Search for $\alpha$-profiles yielding level-crossing}

In the following the described ES will be used to treat
another problem. Whereas in the last subsection we had used 
growth rates resulting from known functions $\alpha(r)$ we 
will now search for a-priori unknown functions. 
Let us assume that we would know
that a cosmic body has a steady 
magnetic field with some field modes, 
say, for $1 \le l \le 4$. If we would know 
(what is, of course, 
not very realistic) that the dynamo mechanism were only 
due to a spherically symmetric $\alpha$ we could ask for 
its radial dependence providing  exactly these four marginal 
modes. This is a typical example our inversion scheme
can be applied for. 

Table 2 gives the functions 
resulting from the inversion, again 
for a number of decreasing regularization parameters $p$, together
with the corresponding values for the first four growth rates. 
We actually give here the resulting power expansions in order to 
allow the reader to validate the results by means
of a standard forward solver for spherically symmetric $\alpha(r)$.

\begin{table}[t]
\caption{Functions $\alpha(r)$ resulting for different 
regularization parameter $p$ 
from the ES inversion procedure for the case
that the modes 
with $1 \le l \le 4$ are demanded to have zero growth rates, and
their computed growth rates.}
\begin{center}
\begin{tabular}{rlrrrr}
\hline
p&$\alpha(r)$ &$\lambda_1$&$\lambda_2$&$\lambda_3$&$
\lambda_4$ \\ \hline
$10^{1}$& $6.84+0.26 \; r^2 -0.003 \; r^3 -0.045 \; r^4$ &
8.89 &5.75&-0.22&-8.89\\
$10^{0}$& $5.70+2.84 \; r^2 +0.27 \; r^3   -0.62 \; r^4$ &
7.52 &5.30&0.19&-7.61\\
$10^{-1}$& $-1.00+14.37 \; r^2   +4.15\; r^3  -4.36 \; r^4$ &
1.58
&3.16&1.29&-3.29\\ 
$10^{-2}$& $-1.81-\-6.02; r^2 +41.52\; r^3   -17.43 \; r^4$
&-1.28 &1.88& 1.34&-1.89\\ 
$10^{-3}$& $5.42-73.11\; r^2 +127.04 \; r^3  
-39.13\; r^4$ &-1.22&1.04&1.11&-1.18\\
$10^{-4}$& $13.89-168.92 \; r^2
+270.60 \; r^3  -91.85\; r^4$ &0.03&-0.32&0.65&-0.32\\ 
$10^{-5}$&
$16.73-269.31\; r^2 +509.56 \; r^3   -237.76\; r^4$ &
-0.03&-0.31&0.80&-0.37\\
$0$& $10.92-99.61 \; r^2 +126.44 \; r^3   -13.17\; r^4$ & 0.16
&0.38&0.90&-0.46\\ \hline \end{tabular}
\end{center}
\end{table}

These functions are depicted also in Fig. 4a.
The dependence of the residual deviation on the curvature 
is again shown in 
Fig. 4b. 
Whereas for large $p$ the 
growth rates
differ significantly from zero, the values for smaller $p$ are 
getting very 
close to zero.
For the best fitting $\alpha(r)$ (at $p=10^{-4}$) we have 
shown in Fig. 4c
the dependence of the growth rates on $C$ for $1 \le l \le 6$. 
We see 
clearly the 
level-crossing at $C \approx 8.5$.

\begin{figure}[t]
\begin{center}
\begin{tabular}{cc}
\epsfxsize=6cm\epsfbox{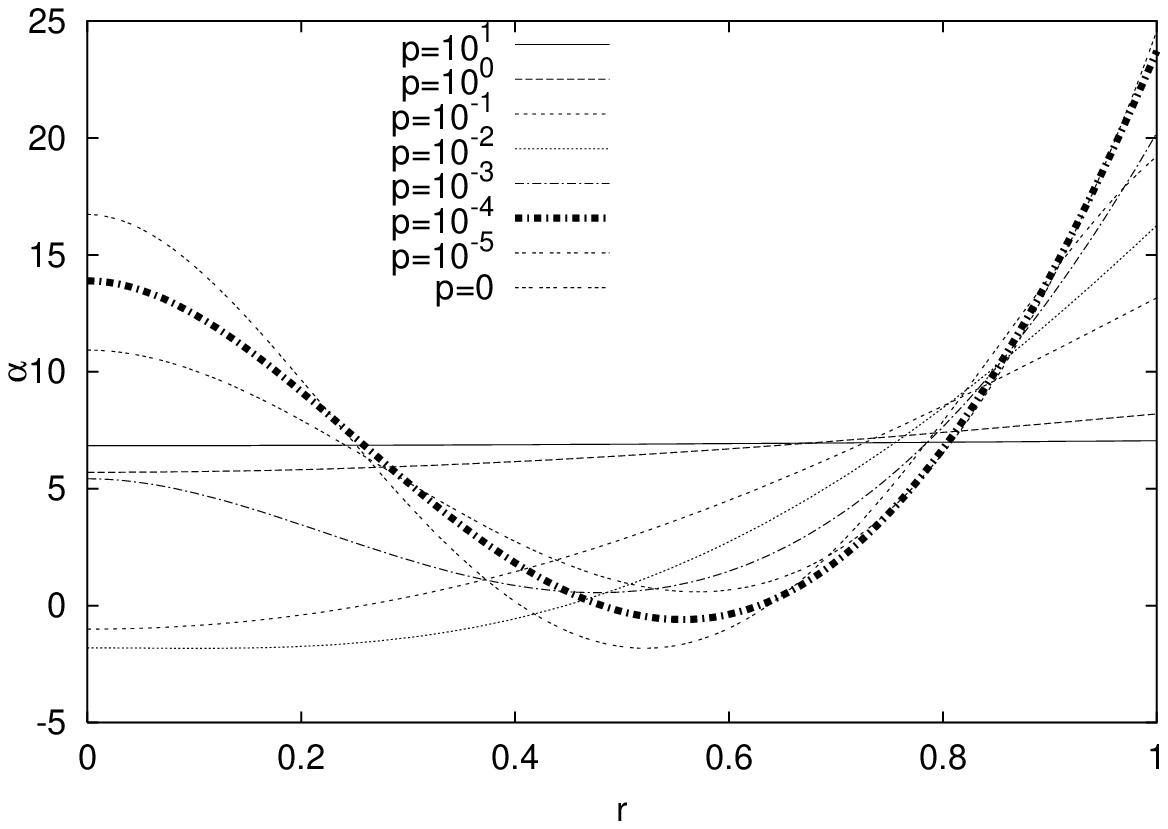} & 
\epsfxsize=6cm\epsfbox{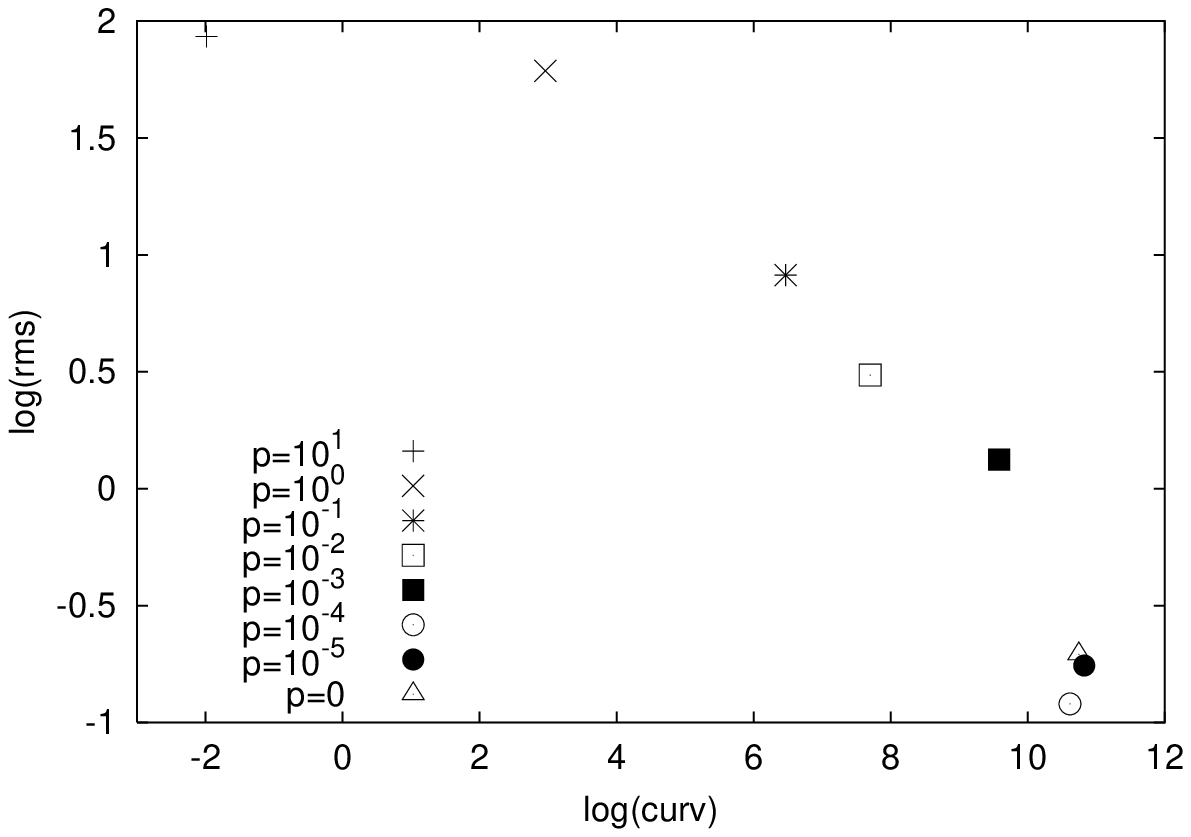}\\
(a)&(b)\\
\epsfxsize=6cm\epsfbox{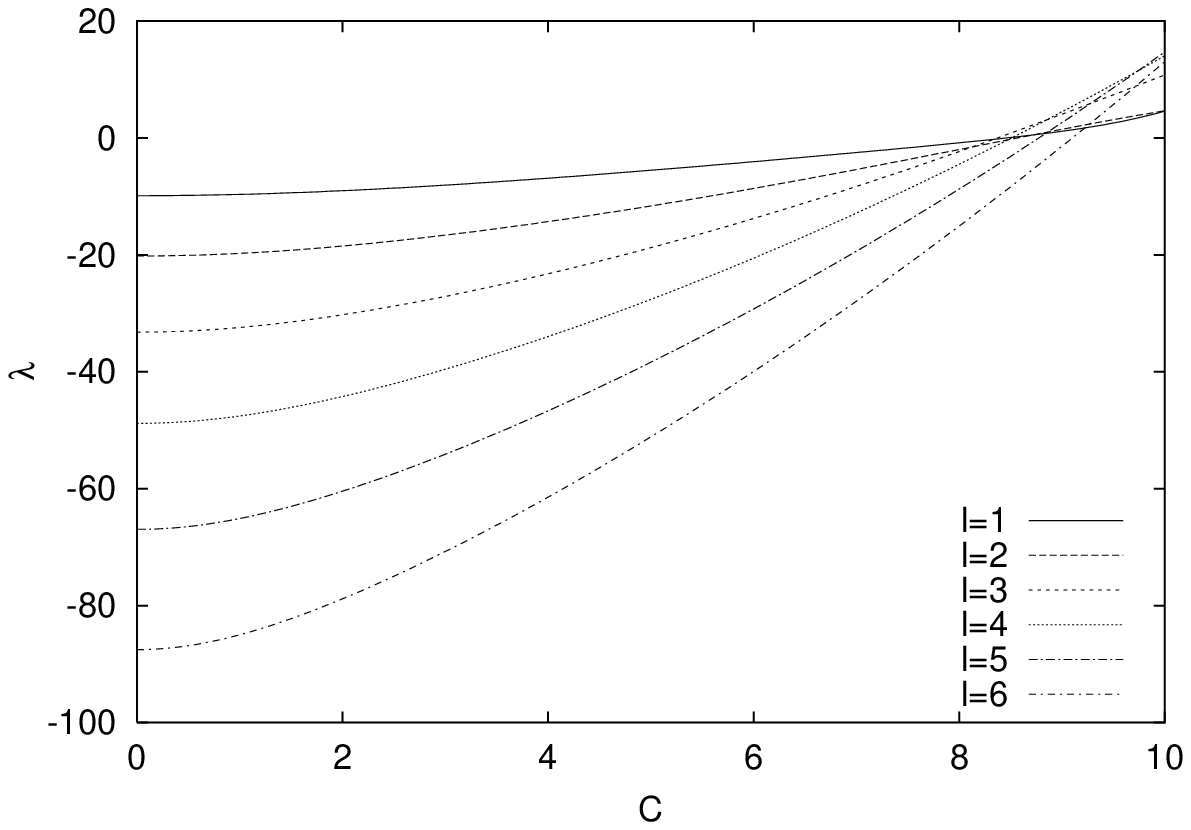}\\
(c)&
\end{tabular}
\end{center}
\caption{Functions $\alpha(r)$ resulting for different 
regularization parameter $p$ from the ES inversion procedure 
for the case
that the modes with $1 \le l \le 4$ are demanded to have 
zero growth rates (a), 
the corresponding dependence of the residual error 
on the curvature (b). For the special regularization
parameter $p=10^{-4}$, the solution 
of the forward problem for the first 6 $l$-modes 
is depicted in (c). 
The ''level-crossing'' at $C \approx 8.5$ is evident.}
\end{figure}

There might also be some relevance of the considered example 
for laboratory dynamos. Let us assume that we were able to build
a spherical dynamo and that the dynamo action of 
its flow were due to
a spherically symmetric $\alpha$. Then, a 
radial dependence of $\alpha$
corresponding to the given curves might lead to a very interesting
behaviour when the back-reaction is taken into 
account. For example it might
happen that the back-reaction changes the 
function $\alpha(r)$ in such a way
that mode switching between the different $l$-modes occurs. 
We would like to suggest further investigation 
of such a model.

\subsection{Optimization}

Let us turn to a second problem which might also be
important for laboratory dynamos. It concerns a kind
of energetic optimization of the dynamo. Of course, the
optimization criterion for a mean field dynamo is not as
obvious as for large scale velocities. We have 
chosen to minimize
the following functional:
\begin{eqnarray}
C'= \left(3 
\int_0^1 \alpha^2(r) r^2 dr \right)^{1/2}
\end{eqnarray}
under the constraint that the mode with $l=1$ becomes marginal.
Of course, the definition (13) is to some extend arbitrary. 
For comparison it is also interesting to consider the
quantity 
\begin{eqnarray}
C''= 3  
\int_0^1  |\alpha(r)| r^2 dr 
\end{eqnarray}
where we used the modulus of $\alpha$ instead of
$\alpha$ itself in order not to get very small
values of $C''$ only due to some cancellation effects 
along the $r$-axis.

To solve this problem, the described ES code had to be 
modified only slightly. 
The resulting functions, together with the corresponding 
values $C'$ and $C''$ are given in Table 3.

We do not claim to have found the function $\alpha(r)$
with the smallest value of $C'$ or $C''$. Remind that 
we have restricted the allowed functions to be
a polynomial expansion up to fourth order. There might be
quite different functions providing even lower values 
of $C'$ and/or $C''$.

Despite some  arbitrariness of the definitions of 
$C'$ and $C''$, it is interesting to note that 
something like the volume averaged ''amount'' of 
$\alpha$ can be decreased significantly compared 
to the well known critical value 4.49 for the 
Krause-Steenbeck dynamo model. 

\begin{table}[t]
\caption{''Energetically''  optimized 
functions $\alpha(r)$ resulting 
from the inversion demanding the mode $l=1$ 
to have zero growth rate. 
$C'$ and $C''$ are defined in the text.}
\begin{center}
\begin{tabular}{rlrr}
\hline
p&$\alpha(r)$ & C'& C''\\ \hline
$10^{-1}$& $6.08-7.93 \; r^2 +3.19 \; r^3   -0.15 \; r^4$ &
3.10     &2.96\\
$10^{-2}$& $8.36-28.24\; r^2 +28.04\; r^3   -8.89 \; r^4$ &
2.57&1.81\\
$10^{-3}$& $11.78-87.25 \; r^2 +124.28 \; r^3   -48.38\; r^4$ &
2.06&0.96\\
$10^{-4}$& $14.24-154.76 \; r^2 +265.25 \; r^3  -124.58 \; r^4$ &
1.88&0.86\\
$10^{-5}$& $14.63-169.03 \; r^2 +298.69 \; r^3   -144.28\; r^4$ &
1.87&0.91\\
$0$& $14.60-168.41 \; r^2 +298.35 \; r^3   -144.54 \; r^4$ &1.88&
0.92\\
\hline
\end{tabular}
\end{center}
\end{table}

\begin{figure}[t]
\begin{center}
\begin{tabular}{cc}
\epsfxsize=8cm\epsfbox{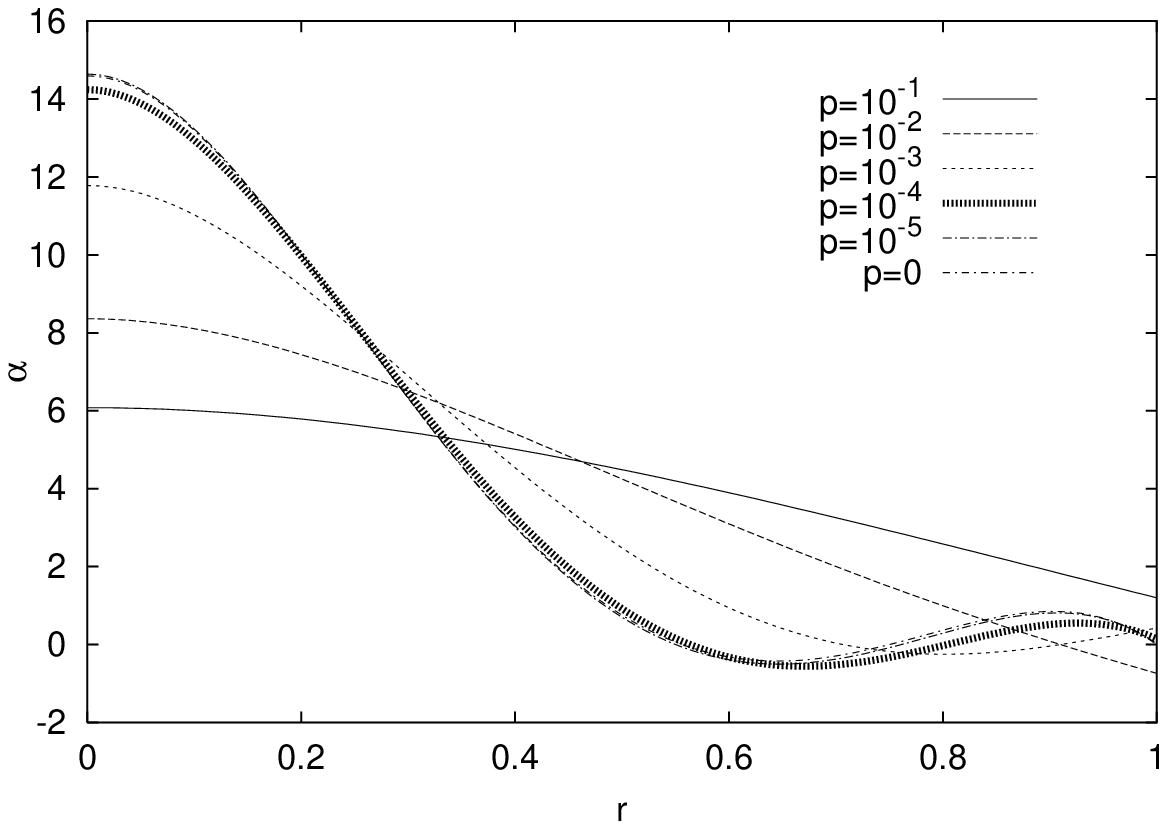}
\end{tabular}
\end{center}
\caption{''Energetically''  optimized 
functions $\alpha(r)$ resulting 
from the inversion demanding the mode 
with $l=1$ to have zero growth rate.}
\end{figure}
  
\section{Conclusions}
In this paper we have tried to reconstruct the 
radial distribution of $\alpha$ from growth rates of 
a few $l$-modes of the magnetic field.

It was shown that some paradigmatic functions
$\alpha(r)$ can indeed be recovered by the used evolutionary 
strategy if a careful regularization of the
trial functions is used. With decreasing 
smoothing effect of the regularization, the 
ES erratically jumps into various local minima and
the resulting functions $\alpha(r)$ can differ 
drastically from the original one. 

Although by our simple numerical means we are not able
to test the isospectrality of different
$\alpha$-profiles in a strict sense, the obvious existence
of different  minima with nearly the same
 quality functions seems
to be more than an accident. A more precise identification
of such isospectral $\alpha$-profiles is highly desirable
and we would like to suggest further analytical and
numerical investigation in this direction. 

On the other side, for any practical application 
(e.g. with measurement errors and only a finite number of
$l$-modes)
it is quite clear that the function 
$\alpha(r)$ can be determined only roughly. 

A unique determination of $\alpha(r)$ would 
probably require knowledge of the spectra for two different 
boundary conditions which is, even for laboratory 
dynamos, 
not very realistic.

An obvious future
task is to generalize the presented 
method to the oscillatory case, possibly in connection 
with varying conductivity (for possible
astrophysical applications of this case, see 
Schubert and Zhang 2000). 
In principle, the developed ES code can be combined with 
any forward solver for more complicated dynamo 
models
in order to apply the method to real cosmic or 
laboratory dynamos. 
For other than spherical geometry it might also 
be useful to combine
the method with an forward solver based on the integral
equation approach (Stefani, Gerbeth, and R\"adler 2000).

\end{document}